\documentclass[a4paper,11pt]{article}
\pdfoutput=1 

\usepackage{jcappub} 
\usepackage{indentfirst}
\usepackage{appendix}
\usepackage{multirow}

\usepackage{array}
\usepackage{enumerate}
\usepackage{graphicx}
\usepackage{dcolumn}
\usepackage{bm}
\usepackage{amssymb}
\usepackage{latexsym}
\usepackage{booktabs}
\usepackage{amsmath}
\usepackage{multirow}
\usepackage{url}

\begin{document}

\title{A path to precision cosmology: synergy between four promising late-universe cosmological probes}

\author[a]{Peng-Ju Wu,}

\author[a]{Yue Shao,}

\author[a]{Shang-Jie Jin}

\author[a,b,c,1]{and Xin Zhang\note{Corresponding author.}}

\affiliation[a]{Key Laboratory of Cosmology and Astrophysics (Liaoning) \& College of Sciences, Northeastern University, Shenyang 110819, China}
\affiliation[b]{Key Laboratory of Data Analytics and Optimization for Smart Industry (Ministry of Education), Northeastern University, Shenyang 110819, China}
\affiliation[c]{National Frontiers Science Center for Industrial Intelligence and Systems Optimization, Northeastern University, Shenyang 110819, China}

\emailAdd{wupengju@stumail.neu.edu.cn, shaoyue@stumail.neu.edu.cn, jinshangjie@stumail.neu.edu.cn, zhangxin@mail.neu.edu.cn}

\abstract{In the next decades, it is necessary to forge new late-universe cosmological probes to precisely measure the Hubble constant and the equation of state of dark energy simultaneously. In this work, we show that the four novel late-universe cosmological probes, 21 cm intensity mapping (IM), fast radio burst (FRB), gravitational wave (GW) standard siren, and strong gravitational lensing (SGL), are expected to be forged into useful tools in solving the Hubble tension and exploring dark energy. We propose that the synergy of them is rather important in cosmology. We simulate the 21\,cm\,IM, FRB, GW, and SGL data based on the hypothetical observations of the Hydrogen Intensity and Real-time Analysis eXperiment, the Square Kilometre Array, the Einstein Telescope, and the Large Synoptic Survey Telescope, respectively. We find that the four probes have different parameter dependencies in cosmological constraints, so any combination of them can break the degeneracies and thus significantly improve the constraint precision. The joint 21\,cm\,IM+FRB+GW+SGL data can provide the constraint errors of $\sigma(\Omega_{\rm m})=0.0022$ and $\sigma(H_0)=0.16\ \rm km\ s^{-1}\ Mpc^{-1}$ in the $\Lambda$CDM model, which meet the standard of precision cosmology, i.e., the constraint precision of parameters is better than 1\%. In addition, the joint data give $\sigma(w)=0.020$ in the $w$CDM model, and $\sigma(w_0)=0.066$ and $\sigma(w_a)=0.25$ in the $w_0w_a$CDM model, which are better than the constraints obtained by the CMB+BAO+SN data. We show that the synergy between the four late-universe cosmological probes has magnificent prospects.}

\maketitle
\section{Introduction}\label{sec1}
Since Edwin Hubble, it has been known that the universe has been expanding. In the late 1990s, type Ia supernovae (SNe Ia) observations revealed that the expansion of the universe is currently accelerating \citep{Riess:1998cb, Perlmutter:1998np}, which means that the gravity becomes a repulsive force on cosmological scales. In physics, usually there are two ways of realizing the cosmic acceleration, i.e., modifying the gravity on large scales or assuming an exotic component having a negative pressure. The latter is known as dark energy.

The cosmological constant $\Lambda$ naturally emerges in general relativity and is considered as the simplest form of dark energy among the possible theoretical hypotheses. The $\Lambda$ cold dark matter ($\Lambda$CDM) model has been viewed as the standard model of cosmology, because it is strongly favoured by the current cosmological observations, in particular the precise measurements of cosmic microwave background (CMB) anisotropies. Recently, however, some cracks appeared in the $\Lambda$CDM model. It has been found that there are tensions between the early- and late-universe measurements; in particular, the Hubble constant ($H_0$) tension is too prominent to be ignored \citep{Riess:2019qba}. Currently, the $H_0$ value measured by the Cepheid-supernova distance ladder is in $4.2\sigma$ tension with that inferred from the Planck CMB observation assuming $\Lambda$CDM \citep{Riess:2020fzl,Aghanim:2018eyx}.

To solve the $H_0$ tension (also known as the ``Hubble tension''), besides searching for new physics in cosmology (see Ref.~\citep{Verde:2019ivm} for a brief review), one should also seek to precisely measure cosmological parameters using only the late-universe observations. Currently, the late-universe observations, such as SNe Ia and the baryonic acoustic oscillations (BAO; here, it refers to those measured from galaxy redshift surveys), cannot tightly constrain cosmological parameters, but can only be used as a supplementary tool to break the parameter degeneracies inherent to the CMB data \citep{Aghanim:2018eyx}. In the next decades, however, some novel late-universe cosmological probes will be greatly developed, in which the most promising ones include, e.g., 21 cm intensity mapping (IM), fast radio burst (FRB), gravitational wave (GW) standard siren, and strong gravitational lensing (SGL).

The 21\,cm\,IM technique provides us with a novel way to measure the large-scale structure (LSS) of the universe. Originated from the photon-baryon plasma prior to recombination, BAO leaves an imprint on the distribution of matter at a characteristic scale of $\sim147$ comoving Mpc \citep{Aghanim:2018eyx}. This scale provides a standard ruler to measure the angular diameter distance $D_{\rm A}(z)$ and the Hubble parameter $H(z)$, and hence allows measurements of the expansion history of the universe (see Ref.~\citep{Weinberg:2013agg} for a review). To measure the BAO signal, one can consider using the 21 cm emission from the neutral hydrogen (\textsc{H\,i}). In the post-reionization era ($z\lesssim6$), most of the \textsc{H\,i} is thought to exist in self-shielded regions embedded in galaxies \citep{Barkana:2006ep}. Therefore, \textsc{H\,i} traces the galaxy distribution, and thus the matter distribution. By mapping the collective \textsc{H\,i} 21 cm emission of many galaxies, one can also obtain the LSS, from which the BAO signals can be extracted. Compared to the traditional galaxy redshift survey method, the 21\,cm\,IM technique is more efficient. We can simply measure the total \textsc{H\,i} intensity within relatively large voxels, instead of having to resolve individual galaxies, which makes it much faster to survey large volumes than galaxy redshift surveys. Therefore, the 21\,cm\,IM surveys could play a crucial role in studying the expansion history of the universe, especially in measuring the equation of state (EoS) of dark energy \cite{Xu:2020uws,Zhang:2019dyq,Zhang:2019ipd,Zhang:2021yof}. The first detection of the 21 cm signal in the IM regime was achieved by Chang et al. in 2010 \citep{Chang:2010jp}. They reported a cross-correlation between 21\,cm\,IM maps and galaxy maps. Since then, several other cross-correlation power spectra between 21\,cm\,IM and galaxies have been detected \citep{Masui:2012zc, Anderson:2017ert, CHIME:2022kvg,Cunnington:2022uzo}. Although the 21\,cm\,IM power spectrum in auto-correlation has not been detected so far, it is believed that a breakthrough can be made in the near future with the vigorous development of 21\,cm\,IM experiments. It is expected that the 21 cm experiments, such as the Square Kilometre Array (SKA) \citep{SKA:2018ckk}, the Baryon acoustic oscillations In Neutral Gas Observations (BINGO) \cite{Battye:2012tg}, the Hydrogen Intensity and Real-time Analysis eXperiment (HIRAX) \cite{Newburgh:2016mwi}, the Canadian Hydrogen Intensity Mapping Experiment (CHIME) \cite{Bandura:2014gwa}, and the full-scale Tianlai cylinder array \citep{Chen:2012xu} will usher in the era of 21-cm cosmology.

FRBs are millisecond-duration radio pulses of unknown progenitors occurring at cosmological distances. They were first discovered by the Parkes telescope in 2007 \citep{Lorimer:2007qn}. Before reaching our radio receivers, the signal from an extragalactic FRB will pass through the host galaxy interstellar medium (ISM), the intergalactic medium (IGM), and the plasma of Milky Way. Therefore, an important characteristic of FRBs is the high dispersion measure (DM), i.e., the long arrival time delay for the low-frequency part of the signal, which is proportional to the number of free electrons existing along the line of sight between the FRB source and the observer \citep{Thornton:2013iua}. Notably, the DM contribution from the IGM, $\rm DM_{IGM}$, can be considered as an indicator of distance to the FRB source, since the intervening electrons increase with increasing distance to the FRB. By measuring the redshift of the host galaxy, one can establish the $\rm DM_{IGM}$--redshift relation, and then place constraints on cosmological parameters \citep{Gao:2014iva}. Compared with the traditional cosmological probes, this probe observes the universe from a unique perspective. For example, due to large DM values of the cosmological FRBs, one can use the localized events to measure the cosmic baryon density $\Omega_{\rm b}$ \citep{Macquart:2020lln, Yang:2022ftm}. In addition, the FRB observations can be used to measure dark energy and the Hubble constant \citep{Zhao:2020ole, Qiu:2021cww,Zhao:2022bpd,Zhao:2022yiv,Wu:2021jyk, Hagstotz:2021jzu, Liu:2022bmn, James:2022dcx}. In recent years, the number of detected FRBs has experienced a dramatic increase, and in 2021, the CHIME/FRB Project released a catalogue of 536 FRBs \citep{CHIMEFRB:2021srp}. So far, the redshifts of 21 FRBs have been determined by identifying their host galaxies \citep{Spitler:2016dmz, Scholz:2016rpt, Chatterjee:2017dqg, Tendulkar:2017vuq, Marcote:2017wan, Bannister:2019iju, Ravi:2019alc, Prochaska:2019, Heintz2020, Bhandari:2020oyb, Marcote:2020ljw, Law:2020cnm, Macquart:2020lln, Bhardwaj:2021xaa, Bhandari:2021pvj, Niu:2021bnl, Ryder:2022qpg, Nimmo:2021ntn, Driessen:2023lxj}. In the era of SKA, about $10^5$--$10^6$ FRBs could be detected per year \citep{Hashimoto:2020dud}, making FRBs promising to be a powerful cosmological probe.

GWs can be utilized as standard sirens \citep{Schutz:1986gp,Holz:2005df}, since the GW waveform directly carries the information of the luminosity distance $D_{\rm L}$ to the GW source. The distance measurement using the standard siren method can obtain absolute luminosity distances (not relative ones), avoiding the distance ladder and calibration process. If the source's redshift can be determined, for example, by identifying an electromagnetic (EM) counterpart of the GW event from the binary coalescence, we can then establish the $D_{\rm L}$--redshift relation, {thereby exploring the expansion history of the universe. In 2017, Abbott et al.~\citep{LIGOScientific:2017adf} applied the multi-messenger observation of GW170817 to cosmological parameter estimation for the first time and obtained the first measurement of the Hubble constant (with 14\% precision) using the standard siren method. The precision of the Hubble constant could reach 2\% using about 50 similar GW standard sirens \citep{Chen:2017rfc}, which is expected to make an arbitration for the Hubble tension. Moreover, GW standard siren observations will be greatly developed in the future. The sensitivity of the Cosmic Explorer (CE)~\citep{LIGOScientific:2016wof} and the Einstein Telescope (ET)~\citep{Punturo:2010zz} will be an order of magnitude improved over the current GW detectors, allowing us to observe GW standard sirens at much higher redshifts \citep{Evans:2021gyd}. Recent works show that CE and ET would play a key role in cosmological constraints \citep{Bian:2021ini,Zhao:2010sz,Cai:2016sby,Zhang:2019ylr,Bachega:2019fki,Belgacem:2019tbw,Wang:2018lun,Zhang:2018byx,Zhang:2019ple,Zhang:2019loq,Li:2019ajo,Jin:2020hmc,Jin:2021pcv,Califano:2022cmo,Jin:2022qnj,Califano:2022syd,Jin:2022tdf,Jin:2023zhi}, especially in breaking the parameter degeneracies of other traditional EM observations \citep{Zhang:2019ple,Zhang:2019loq,Li:2019ajo,Jin:2021pcv,Jin:2020hmc}.} We note that even if some GW events are expected to have no EM counterparts, such as the binary black hole mergers, they could also be used in cosmological fits using the statistical methods \cite{Chen:2017rfc,Feeney:2018mkj,LIGOScientific:2018gmd,Ding:2018zrk,DES:2019ccw,Yu:2020vyy,Wang:2022oou,Song:2022siz}. We expect that GW standard sirens would play an important role in solving the Hubble tension and improving cosmological parameter estimation.

SGL is a rare astronomical phenomenon. As photons from a distant source propagate to detectors on the Earth, their trajectories are deflected by the gravity of intervening mass overdensities, such as galaxies, groups, and clusters. In rare cases, the deflection is sufficiently large to produce multiple images of the light source (see Ref.~\citep{Treu:2010uj} for a review). In the past few decades, many SGL systems have been discovered, giving rise to two important cosmological applications. One is the velocity dispersion (VD) method \citep{Futamase:2000hnr,Grillo:2007iv}, whose key idea is to combine the observations of SGL and stellar dynamics in elliptical galaxies. Specifically, the mass enclosed within the Einstein radius can be derived by measuring the Einstein angle or by measuring the central velocity dispersion of the stellar component. Once the lens mass model is determined, a relation between the Einstein angle and the stellar velocity dispersion can be obtained, and the cosmological parameters can be estimated from this \citep{Grillo:2007iv,Chen:2018jcf,Wang:2019yob,Liu:2021xvc}. The other is the time delay (TD) method \citep{Refsdal:1964nw,Treu:2016ljm}. The TD between multiple images depends on the gravitational potential as well as a ratio of angular diameter distances. The $H_0$ Lenses in COSMOGRAIL's Wellspring (H0LiCOW) collaboration has obtained a measurement of $H_0$ with a 2.4\% precision (for $\Lambda$CDM) utilizing the TDs of six lensed quasars \citep{Wong:2019kwg}, having a $3.1\sigma$ tension with the Planck result \citep{Aghanim:2018eyx} (for an analysis for interacting dark energy, see Ref.~\cite{Wang:2021kxc}). While for the VD measurements, 161 available samples are obtained with well-defined selection criteria using spectroscopic and astrometric observations at present \citep{Chen:2018jcf}. In the Large Synoptic Survey Telescope (LSST) era, more than 8000 SGL systems with well-measured VDs and a few dozens of SGL systems with well-measured TDs could potentially be observed \citep{Oguri:2010ns,Collett:2015roa}, which will play an important role in cosmological constraints \cite{Qi:2022kfg,Wang:2022rvf}.

In this paper, we focus on these four novel late-universe cosmological probes. Although today they have not been truly realized or have poor constraints on cosmology due to limited observational data, they are expected to develop into powerful cosmological tools in the future. Moreover, making use of different physical effects, the four probes are anticipated to have different parameter dependencies and hence can break degeneracies. In this work, we wish to offer an answer to the question of whether the combination of the four promising late-universe probes could provide precise cosmological parameter measurements.

We simulate the 21\,cm\,IM, FRB, GW, and SGL data based on HIRAX, SKA, ET, and LSST, respectively. We give a detailed description of methodology in Sec.~\ref{sec2}, and then we present the forecasted constraints on cosmological parameters and make some discussions in Sec.~\ref{sec3}. Finally, we give our conclusions in Sec.~\ref{sec4}.

\section{Methodology}\label{sec2}
In this paper, we employ the Planck best-fit $\Lambda$CDM model \citep{Aghanim:2018eyx} for the fiducial cosmology, with $H_0=67.3\ \rm km\ s^{-1}\ Mpc^{-1}$, $\Omega_{\Lambda}=0.683$, $\Omega_{\rm m}=0.317$, $\Omega_{\rm b}=0.0495$, $\Omega_{k}=0$, $\sigma_8=0.812$, and $n_{\rm s}=0.965$. Unless otherwise specified, we use $D_{\rm C}(z)$, $D_{\rm A}(z)$, $D_{\rm L}(z)$, $H(z)$, $H_0$, and $E(z)\equiv H(z)/H_0$ to represent comoving distance, angular diameter distance, luminosity distance, Hubble parameter, the Hubble constant, and dimensionless Hubble parameter, respectively. We also define $D_{\nu}(z)=c(1+z)^2/H(z)$, where $c$ is the speed of light.

In our analysis, we first simulate observations and calculate the measurement errors of observables (such as $D_{\rm A}$ and $D_{\rm L}$) for each probe, and then adopt the Markov Chain Monte Carlo (MCMC) method to maximize the likelihood $\mathcal{L}\propto \exp{(-\chi^2/2)}$ to infer the probability distributions of cosmological parameters (such as $H_0$ and $\Omega_{\rm m}$). Note here that, in some simulation processes, we occasionally use the Fisher matrix method to determine the uncertainties of cosmological observables and astrophysical parameters, but in the cosmological parameter estimations, we uniformly use the MCMC method to infer cosmological parameters. {This is because the MCMC method is more accurate than the Fisher matrix method in forecast studies when the situations have some complexities. More specifically, the MCMC method allows for the non-Gaussian distribution of parameters, which is inherited from the non-linearity of the model with respect to the cosmological parameters.}

Next, we will introduce the simulations of the 21\,cm\,IM, FRB, GW, and SGL data in turn, and construct the likelihood function of each probe for the subsequent MCMC analysis.

\subsection{21 cm intensity mapping}\label{sec2.1}

In the IM regime, the spatial location of an observed pixel is given by 2D angular direction ${\boldsymbol{\theta}_{p}}$ and frequency ${\nu}_{p}$ \cite{Bull:2014rha}, i.e.,
\begin{align}
{\boldsymbol{r}}_{\perp} &= D_{\rm C}(z_i)\big({\boldsymbol{\theta}_{p}}-{\boldsymbol{\theta}}_{i}\big),\ r_{\parallel} = D_{\nu}(z_{i})\big(\tilde{\nu}_{p}-\tilde{\nu}_{i}\big),
\end{align}
where the survey has been centered on $\big({\boldsymbol{\theta}}_{i},{\nu}_{i}\big)$, corresponding to a redshift bin centered at $z_{i}$. $\tilde{\nu} \equiv \nu/\nu_{21}$, with $\nu_{21}=1420.4\ \rm MHz$ being the rest-frame frequency of the 21 cm line. We will work in observational coordinates $(\boldsymbol{q}={\boldsymbol{k}}_{\perp}D_{\rm C}, y=k_{\parallel}D_{\nu})$, where $\boldsymbol{k}$ is the wave vector.

The mean \textsc{H\,i} brightness temperature is given by \citep{Hall:2012wd}
\begin{align}
\overline{T}_{\rm b}(z)=188h\Omega_{\textsc{H\,i}}(z)\displaystyle{\frac{(1+z)^2}{E(z)}\,\rm mK},
\end{align}
where $\Omega_{\textsc{H\,i}}(z)$ is the fractional density of \textsc{H\,i} for which we adopt the form shown in Figure 20 of Ref.~\cite{Bull:2014rha}, and $h$ is the dimensionless Hubble constant satisfying $H_0=100h\ \rm km\ s^{-1}\ Mpc^{-1}$. Considering the effect of redshift space distortions \citep{Kaiser:1987qv}, the signal covariance can be written as \citep{Seo:2003pu,Bull:2014rha}
\begin{align}
C^{\rm S}(\boldsymbol{q},y)=\displaystyle{\frac{\overline{T}_{\rm b}^{2}(z_{i})\alpha_{\perp}^2\alpha_{\parallel}}{D_{\rm C}^2D_{\nu}}}\left(b_{\textsc{H\,i}}+f\mu^2\right)^2\exp{\left(-k^2 \mu^2 \sigma_{\rm NL}^{2}\right)}\times P(k),
\end{align}
where $\alpha_{\perp}\equiv D_{\rm A}^{\rm fid}(z)/D_{\rm A}(z)$ and $\alpha_{\parallel}\equiv H(z)/H^{\rm fid}(z)$, with ``fid'' labeling the quantities calculated in the fiducial cosmology. $b_{\textsc{H\,i}}$ is the \textsc{H\,i} bias, and its specific calculation can be found in Ref.~\cite{Xu:2014bya}. $f(z)\approx\Omega_{\rm m}^{\gamma}(z)$ is the linear growth rate with $\gamma=0.545$ for $\Lambda$CDM, and $\mu\equiv k_{\parallel}/k$. $\sigma_{\rm NL}$ is the non-linear dispersion scale, for which we take $7\ \rm{Mpc}$ \citep{Li:2007rpa}, corresponding to a wave vector of $k_{\rm NL}=0.14\ \rm{Mpc^{-1}}$. $P(k)=D^2(z)P(k,z=0)$, with $D(z)$ being the linear growth factor, which is related to $f(z)$ by
\begin{align}
f=\displaystyle{\frac{d\ln{D(a)}}{d\ln{a}}}=-\displaystyle{\frac{1+z}{D(z)}}\displaystyle{\frac{dD(z)}{dz}},
\end{align}
and $P(k,z=0)$ being the matter power spectrum at $z=0$ that can be generated by {\tt CAMB} \citep{Lewis:1999bs}. In this paper, the primordial power spectrum is determined by adopting the Planck best-fit values of $A_{\rm s}$ and $n_{\rm s}$.

Now we turn to instrumental and sky noises as well as effective beams. The noise covariance has the form \citep{Bull:2014rha}
\begin{align}
C^{\rm N}(\boldsymbol{q},y)=\displaystyle{\frac{\sigma_{\rm pix}^2V_{\rm pix}}{D_{\rm C}^2D_{\nu}}} B_{\parallel}^{-1} B_{\perp}^{-2},
\end{align}
where $\sigma_{\rm pix}$ is the pixel noise, $V_{\rm pix}=D_{\rm C}^2{\rm FoV}\times D_{\nu}\delta\nu/\nu_{21}$ is the pixel volume, in which $\rm FoV$ is the field of view of each receiver and $\delta\nu$ is the channel bandwidth. $B_{\parallel}$ and $B_{\perp}$ describe the frequency and angular responses of the instrument, respectively. For HIRAX,
\begin{align}
\sigma_{\rm pix}=\displaystyle{\frac{T_{\rm sys}}{\sqrt{n_{\rm pol}t_{\rm tot}\delta\nu\left(\rm{FoV}/S_{\rm area}\right)}}}
\displaystyle{\frac{\lambda^2}{A_{\rm e}\sqrt{\rm FoV}}}
\displaystyle{\frac{1}{\sqrt{n(\boldsymbol{u})N_{\rm b}}}},
\end{align}
where $T_{\rm sys}$ is the system temperature, $n_{\rm pol}=2$ is the number of polarization channels, $t_{\rm tot}=10,000\, {\rm hr}$ is the observing time, $S_{\rm area}=15,000\, {\rm deg}^2$ is the survey area, $A_{\rm e}=\eta\pi(D_{\rm d}/2)^2$ is the effective collecting area of each receiver, in which $D_{\rm d}=6\,\rm m$ is the diameter of the dish and $\eta=0.7$ is the efficiency factor, $\rm{FoV}\approx(\lambda/D_{\rm d})^2$, $N_{\rm b}=1$ is the number of beams, and $n({\boldsymbol{u}})$ is the baseline density (the detailed calculation can be found in Ref.~\cite{Bull:2014rha}). The system temperature is given by
\begin{align}
T_{\rm sys}=T_{\rm rec} + T_{\rm gal} + T_{\rm CMB},
\end{align}
where $T_{\rm rec}=50\, \rm{K}$ is the receiver noise temperature, $T_{\rm gal}\approx25\ {\rm K}\times(408\ \rm{MHz}/\nu)^{2.75}$ is the contribution from the Milky Way, and $T_{\rm CMB}\approx2.73\ {\rm K}$ is the CMB temperature. Assuming a Gaussian channel bandpass, the effective beam in the parallel direction is given by \citep{Bull:2014rha}
\begin{align}
B_{\parallel}(y)=\exp{\left(-\displaystyle{\frac{(y\delta\nu/\nu_{21})^2}{16\ln{2}}}\right)},
\end{align}
and the transverse effective beam has been accounted for by $n({\boldsymbol{u}})$.

Detection of 21 cm signal is complicated by astrophysical foregrounds that are orders of magnitude brighter than the \textsc{H\,i} signal. Fortunately, these foregrounds have a basically smooth spectral structure and hence one can use some sort of cleaning algorithm to remove them \cite{deOliveira-Costa:2008cxd, Bonaldi:2014zma, Mertens:2017gxw,Hothi:2020dgq,Soares:2021ohm,Ni:2022kxn,Gao:2022xdb}. In this work, we assume that a cleaning algorithm has been applied and the covariance of residual foreground can be modeled as \citep{Bull:2014rha}
\begin{align}
C^{\rm F}(\boldsymbol{q},y)=\varepsilon_{\rm FG}^2\sum_{X}A_{X}\left(\displaystyle{\frac{\ell_p}{2\pi q}}\right)^{n_{X}}\left(\displaystyle{\frac{\nu_p}{\nu_i}}\right)^{m_{X}},
\end{align}
where $\varepsilon_{\rm FG}\in[0,1]$ is a scaling factor, which parameterizes the foreground removal efficiency: $\varepsilon_{\rm FG}=1$ corresponds to no removal and $\varepsilon_{\rm FG}=0$ corresponds to perfect removal. {In this work, we consider an optimistic scenario of $\varepsilon_{\rm FG}=10^{-6}$. It should be emphasized that subtracting the foreground to such a level is extremely challenging. For discussion on the performance of 21\,cm\,IM in the cosmological parameter constraints as the foreground removal efficiency is relatively low, we refer the reader to Ref.~\cite{Wu:2021vfz}.} For a foreground $X$, $A_X$ is the amplitude, $n_X$ and $m_X$ are the angular scale and frequency power-law indices, respectively. These parameters at $\ell_p=1000$ and $\nu_p=130\ \rm MHz$ are given in Ref.~\cite{Santos:2004ju}.

We define the total covariance as $C^{\rm T}=C^{\rm S}+C^{\rm N}+C^{\rm F}$, then the Fisher matrix for a set of parameters $\boldsymbol{p}$ in a redshift bin is given by \citep{Bull:2014rha}
\begin{align}
{F}_{ij}=\displaystyle{\frac{1}{8\pi^2}}V_{\rm bin}\int_{-1}^{1}d\mu\int_{k_{\rm min}}^{k_{\rm max}}k^2dk \displaystyle{\frac{\partial\ln{C^{\rm T}}}{\partial p_{i}}} \displaystyle{\frac{\partial\ln{C^{\rm T}}}{\partial p_{j}}},
\end{align}
where $V_{\rm bin}=S_{\rm area}\Delta\tilde{\nu}D_{\rm C}^2D_{\nu}$ is the survey volume, with $\Delta\tilde{\nu}$ being the dimensionless bandwidth in the redshift bin. In this work, the parameter set $\{\boldsymbol{p}\}$ is selected as $\{D_{\rm A}(z), H(z),$ $[f\sigma_8](z), [b_{\textsc{H\,i}}\sigma_8](z), \sigma_{\rm NL}\}$.
We assume that $b_{\textsc{H\,i}}$ is only redshift dependent, which is appropriate for large scales, so we impose a non-linear cut-off at $k_{\rm max}=k_{\rm NL}(1+z)^{2/(2+n_{\rm s})}$ (with fixed $n_{\rm s}$ by Planck observation) \citep{Smith:2002dz}. Moreover, the largest scale the survey can probe corresponds to a wave vector $k_{\rm min}=2\pi V_{\rm bin}^{-1/3}$.

\begin{figure*}[!htbp]
\includegraphics[scale=0.35]{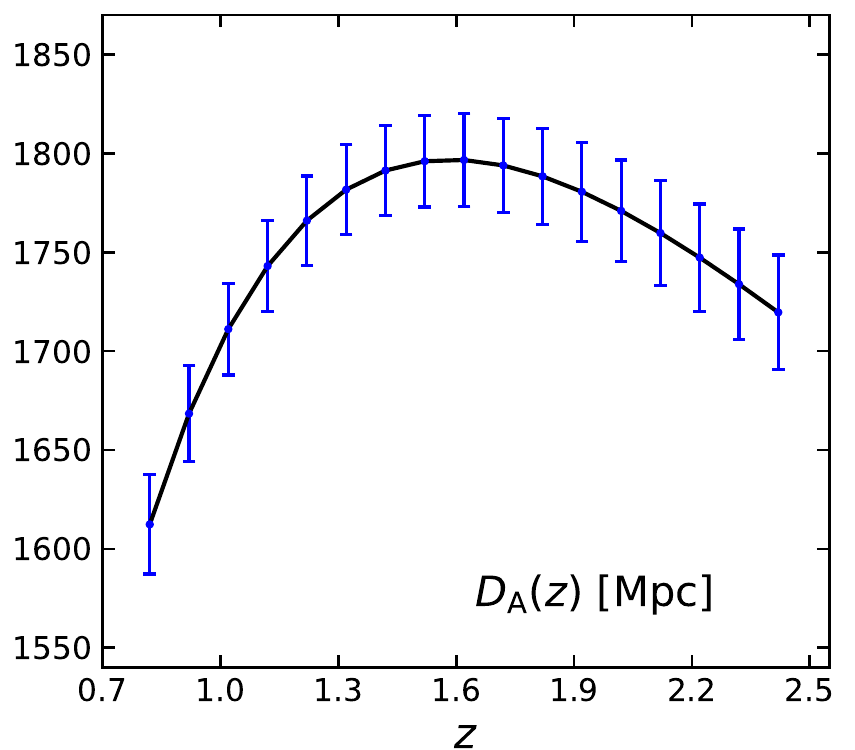}
\includegraphics[scale=0.35]{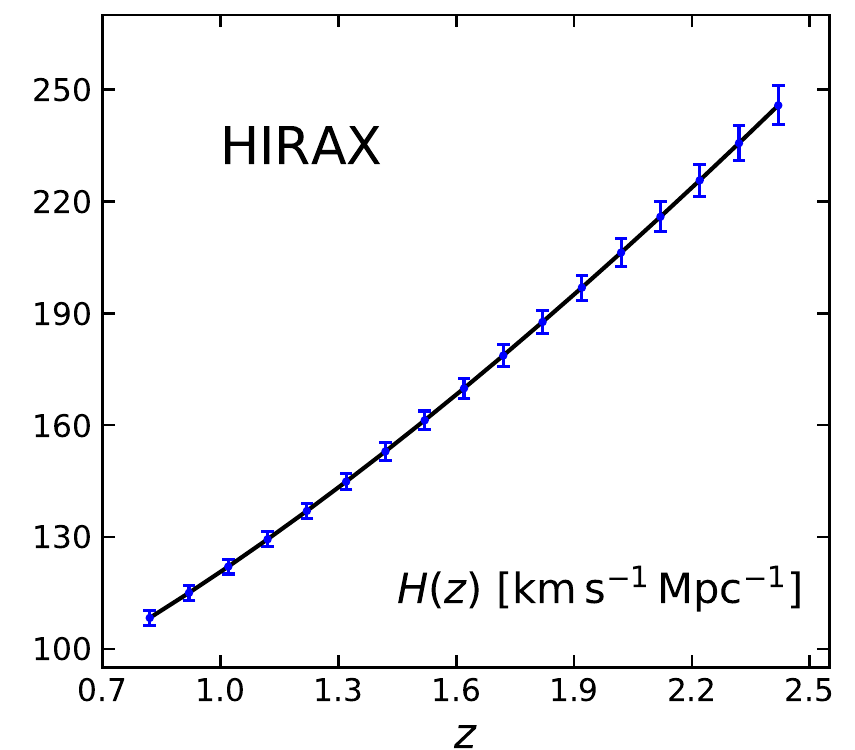}
\includegraphics[scale=0.35]{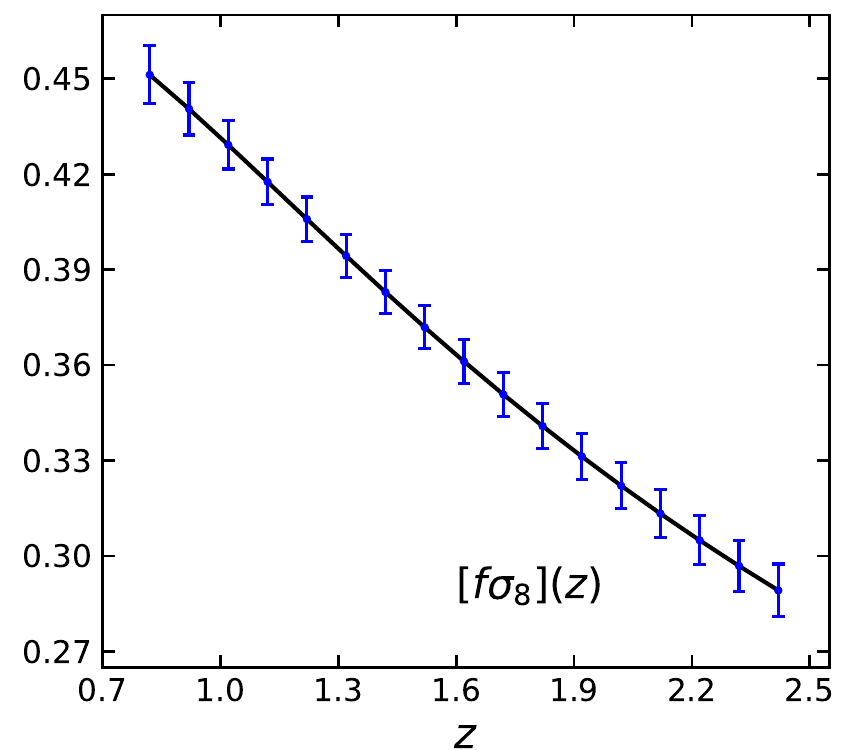}
\caption{Measurement errors on $D_{\rm A}(z)$ (left panel), $H(z)$ (central panel), and $[f\sigma_8](z)$ (right panel) of HIRAX.}
\centering
\label{21cmIM}
\end{figure*}

\begin{figure}[!htbp]
\includegraphics[scale=0.55]{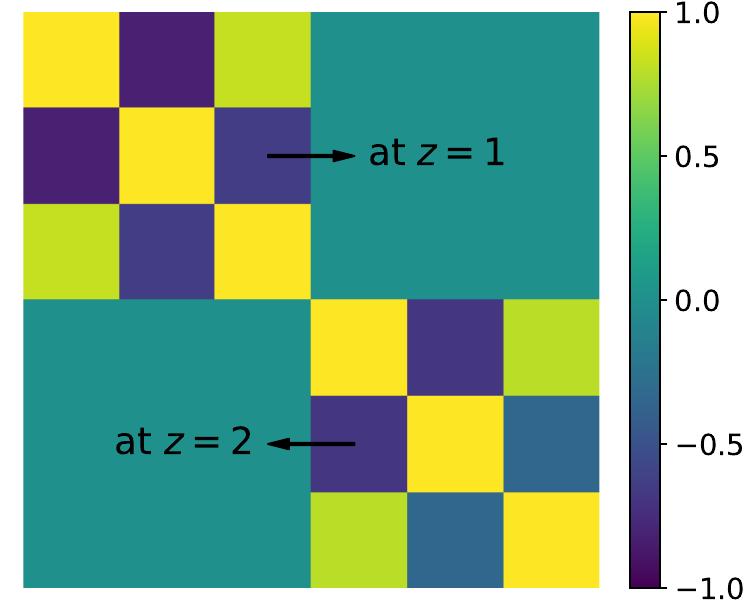}
\centering
\caption{The normalized covariance matrix concerning $\{D_{\rm A}(z), H(z), [f\sigma_8](z)\}$ at $z=1$ and $2$, respectively.}
\label{Cov}
\end{figure}

We simulate the measurement of 21\,cm\,IM power spectrum and calculate the Fisher matrix for $\boldsymbol{p}$ in each redshift bin. The bin width is 0.1, and there are 17 bins in total. Note that we marginalize $[b_{\textsc{H\,i}}\sigma_8](z)$ and $\sigma_{\rm NL}$, and construct the inverse covariance matrix for $\{D_{\rm A}(z), H(z), [f\sigma_8](z)\}$ (which equals to the Fisher matrix) for the next step of cosmological parameter estimation. As mentioned earlier, we use MCMC to give constraints on cosmological parameters $\boldsymbol{\xi}$. The $\chi^2$ function for a redshift bin is given by
\begin{align}
\chi^2(\boldsymbol{\xi}) = \sum_{ij}{x}_i{F}_{ij}{x}_j,
\end{align}
where
\begin{align}
\boldsymbol{x}=(H^{\rm th}(\boldsymbol{\xi})-H^{\rm obs},\, D&_{{\rm A}}^{\rm th}(\boldsymbol{\xi})-D_{{\rm A}}^{\rm obs},\, {[f\sigma_8]}^{\rm th}(\boldsymbol{\xi})-{[f\sigma_8]}^{\rm obs}).
\end{align}
Here $\boldsymbol{\xi}$ denotes the set of cosmological parameters and the Fisher matrix $F_{ij}$ serves as the inverse covariance matrix concerning $\{D_{\rm A}(z), H(z), [f\sigma_8](z)\}$. The total $\chi^2$ function of 21\,cm\,IM, $\chi^2_{\rm 21\,cm\,IM}$, is the sum of the $\chi^2$ functions of all redshift bins. The mock data are shown in Fig.~\ref{21cmIM}. {To illustrate the correlations between observables, we show two representative normalized covariance matrices, at redshifts 1 and 2, respectively, in Fig.~\ref{Cov}. Note that we only consider the correlations between observables at the same redshift, and the correlations between different redshifts are not considered.}

\subsection{Fast radio burst}\label{sec2.2}
In order to mock future detectable FRBs, we need to assume a redshift distribution of FRBs. In this work, we assume that the comoving number density of FRB sources is proportional to the cosmic star formation history (SFH) \citep{Hopkins:2006bw} (see also Refs.~\cite{Hashimoto:2022llm,Li:2023zro}), then the redshift distribution of FRBs is given by \citep{Li:2019klc}
\begin{align}
N_{\rm SFH}(z)=\mathcal{N}_{\rm SFH}\displaystyle{\frac{\dot{\rho}_*(z)D_{\rm C}^2(z)}{H(z)(1+z)}e^{{-D_{\rm L}^2(z)}/{[2D_{\rm L}^2(z_{\rm cut})]}}},
\end{align}
where $\mathcal{N}_{\rm SFH}$ is a normalization factor. $z_{\rm cut} = 1$ is a cutoff, which characterizes the decrease of detected FRBs beyond it due to the instrumental signal-to-noise threshold effect. The density evolution can be parameterized as \citep{Caleb:2015uuk}
\begin{align}
\dot{\rho}_*(z)=a_1\displaystyle{\frac{a_2+a_3z}{1+(z/a_4)^{a_5}}},
\end{align}
with $a_1=0.7$, $a_2=0.017$, $a_3=0.13$, $a_4=3.3$, and $a_5=5.3$.

When considering the observed DM of an FRB, the contributions from various ionized regions along the line of sight can be separated as \citep{Thornton:2013iua, Deng:2013aga}
\begin{align}
\label{DM}
{\rm DM}_{\rm obs}={\rm DM}_{\rm host}+{\rm DM}_{\rm IGM}+{\rm DM}_{\rm MW}.
\end{align}
Here, ${\rm DM}_{\rm host}$ refers to the contribution from the host galaxy interstellar medium and ${\rm DM}_{\rm MW}$ represents the contribution by the plasma of the Milky Way. The ${\rm DM}_{\rm IGM}$ term relates to cosmology, since the intervening electrons increase with increasing distance to the FRB, and its average value can be expressed as
\begin{align}
\label{FRB-DM-IGM}
\overline{{\rm DM}}_{\rm IGM}(z)=\displaystyle{\frac{3cH_0\Omega_{\rm b}}{8\pi Gm_{\rm p}}}\int_{0}^{z}\displaystyle{\frac{(1+z')f_{\rm IGM}(z')\chi (z')dz'}{E(z')}},
\end{align}
where
\begin{align}
\chi(z)=Y_{\rm H}\,\chi_{\rm e,H}(z)+\displaystyle{\frac{1}{2}}Y_{\rm He}\,\chi_{\rm e,He}(z).
\end{align}
In this expression, $\Omega_{\rm b}$ is the present-day baryon density parameter and $f_{\rm IGM}$ is the baryon mass fraction in the IGM for which we adopt $0.053z+0.82$ at $z\leq1.5$ and $0.9$ at $z>1.5$ \citep{Zhou:2014yta}. $Y_{\rm H}=3/4$ and $Y_{\rm He}=1/4$ are the mass fractions of hydrogen and helium, respectively, and $\chi_{\rm e,H}$ and $\chi_{\rm e,He}$ are the ionization fractions for H and He, respectively. We assume $\chi_{\rm e,H}=\chi_{\rm e,He}=1$, which are reasonable at $z<3$ because the IGM is almost fully ionized.

Now we turn to errors in ${\rm DM}_{\rm IGM}$ measurement. From Eq.~(\ref{DM}), we can infer ${\rm DM}_{\rm IGM}$ if ${\rm DM}_{\rm host}$, ${\rm DM}_{\rm MW}$ and ${\rm DM}_{\rm obs}$ could be determined. Thus, we can calculate the total uncertainty of ${\rm DM}_{\rm IGM}$ using the expression
\begin{align}
\sigma_{{\rm DM}_{\rm IGM}}=\left[\left(\displaystyle{\frac{\sigma_{\rm host}}{1+z}}\right)^2 + \sigma_{\rm IGM}^2 + \sigma_{\rm MW}^2 + \sigma_{\rm obs}^2 \right]^{1/2}.
\end{align}
It is difficult to estimate uncertainty of ${\rm DM}_{\rm host}$, because it depends strongly on host galaxy type and local environment. Here we take $\sigma_{\rm host}=30\, \rm {pc\,cm^{-3}}$. The factor $1+z$ accounts for cosmological time dilation for a source at redshift $z$. The uncertainty $\sigma_{\rm IGM}$ describes the deviation of an individual event from the mean $\rm DM_{IGM}$, due to the inhomogeneity of the baryon matter in the IGM, and we adopt the form \citep{Jaroszynski:2020kqy}
\begin{align}
\sigma_{\rm IGM}\simeq 0.2{\rm DM}_{\rm IGM}\,z^{-1/2}.
\end{align}
According to the Australia Telescope National Facility pulsar catalogue \citep{Manchester:2004bp}, the average uncertainty of $\rm DM_{MW}$ for the sources at high Galactic latitude is about $10\, \rm {pc\,cm^{-3}}$. The observational uncertainty $\sigma_{\rm obs}=1.5\, \rm {pc\,cm^{-3}}$ is adopted from the average value of the released data \citep{Petroff:2016tcr}.

For cosmological studies, we need to estimate the FRB event rate. Hashimoto et al. \citep{Hashimoto:2020dud} pointed out that $\sim10^4$ FRB events per day could be detected by the upcoming SKA. Assuming that only 1\% of the detected FRBs can be sufficiently localized to confirm their host galaxies, there are still $\sim100$ FRBs available per day for cosmological constraints. In this work, we consider an optimistic scenario of $N_{\rm FRB} = 100,000$ for a few years of observation (see also Ref.~\cite{Zhao:2022bpd}). For the performance of $10,000$ FRBs (a relatively conservative scenario) in cosmological constraints, we refer the reader to Refs.~\citep{Zhao:2020ole,Qiu:2021cww}.

Assuming that the FRB events are independent of each other, the $\chi^2$ function of FRB can be written as
\begin{align}
\chi^2_{\rm FRB}(\boldsymbol{\xi})=\sum_{i=1}^{100,000}\left(\displaystyle{\frac{{\rm DM}_{{\rm IGM},i}^{\rm th}(\boldsymbol{\xi})-{\rm DM}_{{\rm IGM},i}^{\rm obs}}{\sigma({\rm DM}_{{\rm IGM},i})}}\right)^2.
\end{align}
We note that Reischke et al.~\citep{Reischke:2023blu} recently explored the covariance matrix of DMs of FRBs induced by the LSS of the universe, and they found that the covariance needs to be taken into account for unbiased inference when future samples contain a few hundred FRBs. In this work, we focus on the synergy of multiple probes, and the covariance is not considered.

\begin{figure}[!htbp]
\includegraphics[scale=0.45]{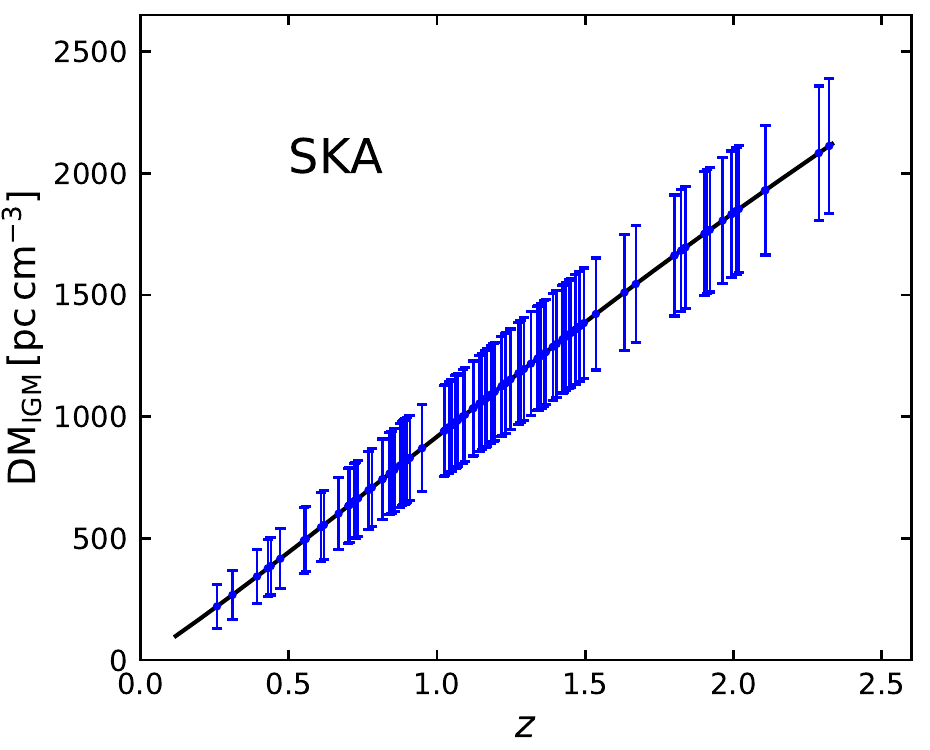}
\includegraphics[scale=0.45]{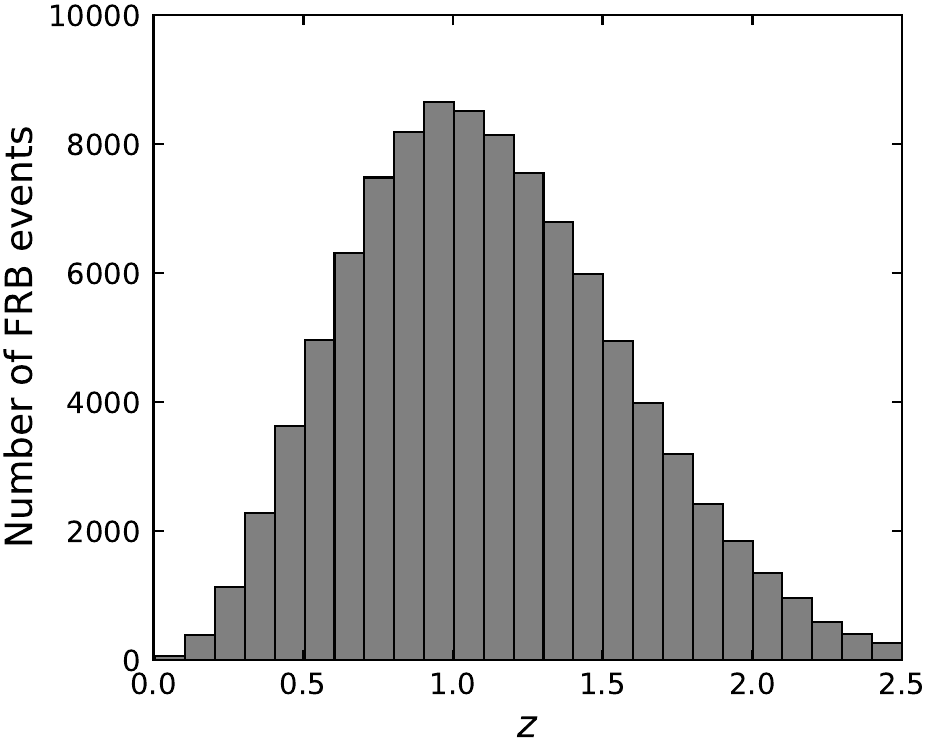}
\centering
\caption{The simulated FRB data based on SKA. Left panel: the 100 representative FRB events. Right panel: the redshift distribution of FRB event number.}
\label{FRB}
\end{figure}

The simulated FRB events are shown in Fig.~\ref{FRB}. In the left panel, we present the 100 representative FRB events, and in the right panel, we show the  redshift distribution of FRB event number.

\subsection{Gravitational wave}\label{sec2.3}
In this paper, we consider that all GW standard sirens detected by ET are provided by the BNS mergers. For the redshift distribution of BNSs, we employ the form \citep{Zhao:2010sz,Cai:2017plb,Zhang:2018byx,Zhang:2019ple,Zhang:2019loq,Jin:2022tdf}
\begin{align}
P(z)\propto \displaystyle{\frac{4\pi D_{\rm C}^2(z)R(z)}{H(z)(1+z)}},
\end{align}
where $R(z)$ is the time evolution of the burst rate,

\begin{eqnarray}
R(z) =
\begin{cases}
1+2z,                                 &  z\leq 1, \\
\displaystyle{\frac{3}{4}}(5-z),      &  1 < z < 5,\\
0,                                    &  z \geq 5.
\end{cases}
\end{eqnarray}

In the transverse-traceless gauge, the GW signal $h(t)$ is the linear combination of the two polarization components (i.e., $h_+$ and $h_{\times}$),
\begin{align}
h(t)=F_+(\theta,\phi,\psi)h_+(t) + F_{\times}(\theta,\phi,\psi)h_{\times}(t),
\end{align}
where $F_+$ and $F_{\times}$ are the antenna pattern functions, $(\theta,\phi)$ are the location angles of the source in the detector frame, and $\psi$ is the polarization angle. The antenna pattern functions of ET are \cite{Zhao:2010sz}
\begin{align}
F_+^{(1)}(\theta, \phi, \psi)=&\frac{{\sqrt 3 }}{2}\bigg[\frac{1}{2}\big(1 + {\cos ^2}\theta \big)\cos(2\phi)\cos(2\psi) - \cos \theta \sin (2\phi)\sin(2\psi)\bigg],\nonumber\\
F_\times^{(1)}(\theta, \phi, \psi)=&\frac{{\sqrt 3 }}{2}\bigg[\frac{1}{2}\big(1 + {\cos^2}\theta \big)\cos(2\phi) \sin(2\psi)+\cos\theta \sin(2\phi)\cos (2\psi )\bigg],\nonumber\\
F_{+,\times}^{(2)}(\theta, \phi, \psi)=&F_{+,\times}^{(1)}(\theta, \phi+2\pi/3, \psi),\nonumber\\
F_{+,\times}^{(3)}(\theta, \phi, \psi)=&F_{+,\times}^{(1)}(\theta, \phi+4\pi/3, \psi).
\end{align}
We use the post-Newtonian approximation (to 3.5 order) to calculate the waveform \cite{Sathyaprakash:2009xs,Zhao:2017cbb}. The Fourier transform $\mathcal{H}(f)$ of the time-domain waveform $h(t)$ is given by
\begin{align}
\mathcal{H}(f)&=\mathcal{A}f^{-7/6}\exp {\left\{i\big[2 \pi f t_{\rm c}-\pi / 4+2 \psi(f / 2)-\varphi_{(2,0)}\big]\right\}},
\end{align}
where the Fourier amplitude $\mathcal{A}$ is calculated by
\begin{align}
\mathcal{A}=\frac{1}{D_{\rm L}}\sqrt{F_+^2\big(1+\cos^2\iota\big)^2+4F_\times^2\cos^2\iota} \times \sqrt{5\pi/96}\,\pi^{-7/6}\mathcal{M}_{\rm c}^{5/6},
\end{align}
and the functions $\psi(f)$ and $\varphi_{(2,0)}$ can refer to Ref. \cite{Sathyaprakash:2009xs}. Here, $t_{\rm c}$ is the epoch of the merger, $\iota$ is the inclination angle between the binary's orbital angular momentum and the line of sight, $\mathcal{M}_{\rm c}=(1+z)\eta^{3/5}M$ is the observed chirp mass, $M=m_1+m_2$ is the total mass of binary system with the component masses $m_1$ and $m_2$, and $\eta=m_1 m_2/M^2$ is the symmetric mass ratio.

In our simulation, only GW events with signal-to-noise ratio (SNR) greater than 8 are selected. The combined SNR for the three detectors of ET is given by
\begin{equation}
\rho=\sqrt{\sum\limits_{n=1}^{3}\left(\mathcal{H}^{(n)}|\mathcal{H}^{(n)}\right)},
\end{equation}
where the inner product is defined as
\begin{equation}
\left({a|b}\right)=4\int_{f_{\rm lower}}^{f_{\rm upper}}\frac{a(f) b^\ast(f)+ a^\ast(f) b(f)}{2}\frac{df}{S_{\rm n}(f)}.\label{snr2}
\end{equation}
Here, $f_{\rm lower}=1\ \rm{Hz}$ is the lower cutoff frequency and $f_{\rm upper}=2f_{\rm LSO}$ is the upper cutoff frequency, where $f_{\rm LSO}=1/{(6^{3/2}2\pi M_{\rm obs})}$ is the orbit frequency at the last stable orbit with $M_{\rm obs}=(1+z)M$ being the observed total mass. $S_{\rm n}(f)$ is the one-side noise power spectral density for ET \cite{ETcurve-web}. Following the estimate in Refs.~\citep{Sathyaprakash:2009xs,Cai:2017plb}, we simulate 1000 GW standard sirens generated by BNS mergers during a 10-year operation of ET.

The measurement errors of $D_{\rm L}$ consist of the instrumental error $\sigma_{D_{\rm L}}^{\rm inst}$, the weak-lensing error $\sigma_{D_{\rm L}}^{\rm lens}$, and the peculiar velocity error $\sigma_{D_{\rm L}}^{\rm pv}$, i.e.,
\begin{align}
\sigma_{D_{\rm L}}=\left[(\sigma_{D_{\rm L}}^{\rm inst})^2+ (\sigma_{D_{\rm L}}^{\rm lens})^2+ (\sigma_{D_{\rm L}}^{\rm pv})^2\right]^{1/2}.
\end{align}
We use the Fisher matrix to calculate $\sigma_{D_{\rm L}}^{\rm inst}$. For ET, the Fisher matrix for a parameter set $\boldsymbol{p}$ is given by
\begin{align}
F_{ij}=\sum_{n=1}^3\left(\frac{\partial {\mathcal{H}}^{(n)}}{\partial p_i}\bigg|\frac{\partial{\mathcal{H}}^{(n)}}{\partial p_j}\right).
\end{align}
We choose the parameter set $\boldsymbol{p}$ as $\{$$D_{\rm L}$, $M_{\rm c}$, $\eta$, $\theta$, $\phi$, $\psi$, $\iota$, $t_{\rm c}$, $\psi_{\rm c}$$\}$, in which $\psi_{\rm c}$ is the coalescence phase, and then
\begin{align}
\sigma_{D_{\rm L}}^{\rm inst} =\sqrt{({F}^{-1})_{11}}\ .
\end{align}
For the error caused by the weak lensing, we adopt \citep{Hirata:2010ba,Tamanini:2016zlh,Speri:2020hwc}
\begin{align}
\sigma_{D_{\rm L}}^{\rm lens}(z)=D_{\rm L}(z)\times 0.066\bigg[\frac{1-(1+z)^{-0.25}}{0.25}\bigg]^{1.8}F_{\rm {d}},
\end{align}
where $F_{\rm d}$ is a delensing factor with the form \citep{Speri:2020hwc}
\begin{align}
F_{\rm {d}}(z)=1-\frac{0.3}{\pi / 2} \arctan \left(z / 0.073\right).
\end{align}
The error caused by the peculiar velocity of the GW source is given by \cite{Kocsis:2005vv}
\begin{align}
\sigma_{D_{\rm L}}^{\rm pv}(z)=D_{\rm L}(z)\times \bigg[ 1+ \frac{c(1+z)^2}{H(z)D_{\rm L}(z)}\bigg]\frac{\sqrt{\langle v^2\rangle}}{c},
\end{align}
where $\sqrt{\langle v^2\rangle}=500\ {\rm km\ s^{-1}}$ is the rms peculiar velocity.

In the process of simulation, we set $t_{\rm c}=0$ for simplicity. Moreover, we assume that the redshifts of the GW sources can be determined by measuring their EM counterparts, such as the short $\gamma$-ray bursts (SGRBs). Notably, the $\gamma$-ray emission is supposed to be confined to a cone with an opening angle as large as $40^{\circ}$, corresponding to inclination angle $\iota=20^\circ$ \cite{li2015extracting}. Finally, for each GW event, the parameters we sample in the ranges of $\theta\in[0, \pi]$, $\phi\in[0, 2\pi]$, $m_1\in[1, 2]\,M_{\odot}$, $m_2\in[1, 2]\,M_{\odot}$, $\iota\in[0, \pi/9]$, $\psi\in[0, 2\pi]$, and $\psi_{\rm c}\in[0, 2\pi]$, respectively, where $M_{\odot}$ is the solar mass.

Since the GW events are independent of each other, the $\chi^2$ function of GW can be written as
\begin{align}
\chi^2_{\rm GW}(\boldsymbol{\xi})=\sum_{i=1}^{1000}\left(\displaystyle{\frac{D_{{\rm L},i}^{\rm th}(\boldsymbol{\xi})-D_{{\rm L},i}^{\rm obs}}{\sigma (D_{{\rm L},i})}}\right)^2.
\end{align}
Here $\boldsymbol{\xi}$ denotes a set of cosmological parameters.

\begin{figure}[!htbp]
\includegraphics[scale=0.45]{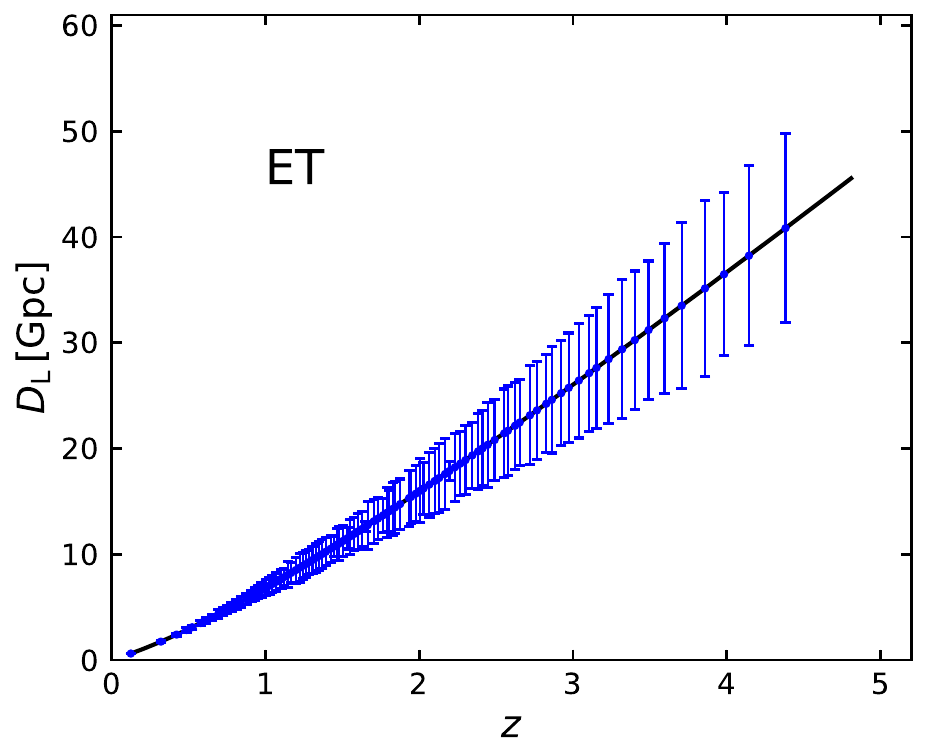}
\includegraphics[scale=0.45]{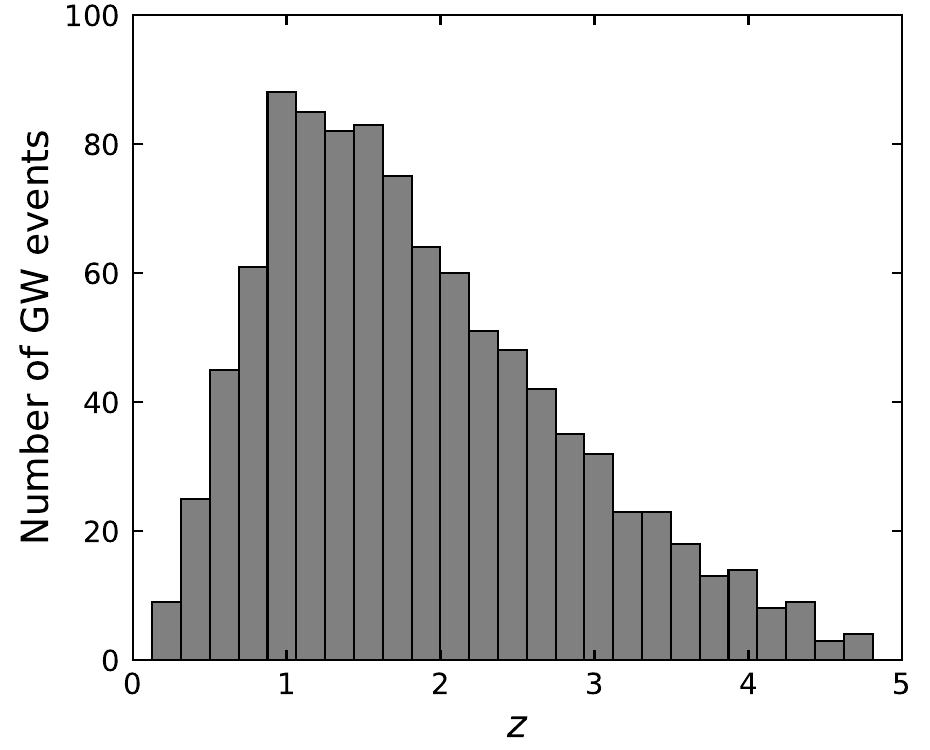}
\centering
\caption{The simulated GW data based on ET. Left panel: the 100 representative GW events. Right panel: the redshift distribution of GW event number.}
\label{GW}
\end{figure}

The simulated GW events are shown in Fig.~\ref{GW}. In the left panel, we present the 100 representative GW events, and in the right panel, we show the redshift distribution of GW event number.

\subsection{Strong gravitational lensing}\label{sec2.4}
In this work, we consider only the galaxy-scale lenses, which dominate the lens abundance \citep{Oguri:2010ns}. One cosmological application of SGL is to combine the observations of SGL and stellar dynamics in elliptical galaxies. The main idea is that the gravitational mass $M_{\rm grl}^{\rm E}$ and dynamical mass $M_{\rm dyn}^{\rm E}$ enclosed within a cylinder of the Einstein radius $R_{\rm E}$ should be equivalent, i.e.,
\begin{align}
\label{eq:SGl}
M_{\rm grl}^{\rm E}=M_{\rm dyn}^{\rm E}.
\end{align}
The gravitational mass $M_{\rm grl}^{\rm E}$ is given by \citep{Chen:2018jcf}
\begin{align}
M_{\rm grl}^{\rm E}=\displaystyle{\frac{c^2}{4G}}\displaystyle{\frac{D_{\rm l}D_{\rm s}}{D_{\rm ls}}}\theta_{\rm E}^2,
\end{align}
where $D_{\rm l}$, $D_{\rm s}$, and $D_{\rm ls}$ are the angular diameter distances between observer and lens, between observer and source, and between lens and source, respectively. $\theta_{\rm E}\equiv R_{\rm E}/D_{\rm l}$ is the Einstein angle. By observing the VD of the lens galaxy and assuming a lens mass model, one can infer the dynamical mass $M_{\rm dyn}^{\rm E}$.

We choose a general mass model based on power-law density profiles for the lens galaxies \citep{Koopmans:2005ig}
\begin{eqnarray}
\begin{cases}
\rho(r) = \rho_0(r/r_0)^{-\gamma}   \\
v(r) = v_0(r/r_0)^{-\delta}   \\
\beta(r) = 1 - (\sigma_{\theta}/\sigma_r)^2,
\end{cases}
\end{eqnarray}
where $\rho(r)$ is the total (i.e., luminous plus dark matter) mass density distribution, $v(r)$ denotes the density distribution of luminous mass, and $\gamma$ and $\delta$ are power-law indices. $\beta(r)$ characterizes the anisotropy of the stellar velocity dispersion, and $\sigma_{\theta}$ and $\sigma_r$ are the tangential and radial components of the velocity dispersion, respectively. Then the mass contained within a sphere with radius $r$ can be written as \citep{Schwab:2009nz}
\begin{align}
\label{eq:Mr}
M(r)=\displaystyle{\frac{2}{\sqrt{\pi}}}\displaystyle{\frac{\Gamma(\gamma/2)}{\Gamma[(\gamma-1)/2]}}\left(\displaystyle{\frac{r}{R_{\rm E}}}\right)^{3-\gamma}M_{\rm dyn}^{\rm E},
\end{align}
where $\Gamma(x)$ is Euler's Gamma function. The radial distance velocity dispersion of the luminous mass could be expressed as
\begin{align}
\label{eq:sigmar}
\sigma_r^2(r)=\displaystyle{\frac{G\int_{r}^{\infty}{\rm d}r'r'^{2\beta-2}v(r')M(r')}{r^{2\beta}v(r)}}.
\end{align}
By substituting Eq.~(\ref{eq:Mr}) into Eq.~(\ref{eq:sigmar}), one reads \citep{Chen:2018jcf}
\begin{align}
\sigma_r^2(r)=\displaystyle{\frac{2}{\sqrt{\pi}}}\displaystyle{\frac{GM_{\rm dyn}^{\rm E}}{R_{\rm E}}}\displaystyle{\frac{1}{\xi-2\beta}}\displaystyle{\frac{\Gamma(\gamma/2)}{\Gamma[(\gamma-1)/2]}}\left(\displaystyle{\frac{r}{R_{\rm E}}}\right)^{2-\gamma},
\end{align}
where $\xi=\gamma+\delta-2$, and $\beta$ is assumed to be independent of the radius $r$.

In practice, what we measure is the luminosity-weighted average of the line-of-sight velocity dispersion of the lens galaxy inside certain apertures $\theta_{\rm ap}$. Moreover, all velocity dispersions $\sigma_{\rm ap}$ measured within $\theta_{\rm ap}$ should be normalized to the one within typical physical aperture $\theta_{\rm eff}/2$, with $\theta_{\rm eff}$ being the effective angular radius of the lens galaxy. The theoretical value of the velocity dispersion within $\theta_{\rm eff}/2$ is given by \citep{Chen:2018jcf}
\begin{small}
\begin{align}
\sigma_0=\sqrt{\displaystyle{\frac{c^2}{2\sqrt{\pi}}}\displaystyle{\frac{D_{\rm s}}{D_{\rm ls}}}\theta_{\rm E}\displaystyle{\frac{3-\delta}{(\xi-2\beta)(3-\xi)}}F\left(\displaystyle{\frac{\theta_{\rm eff}}{2\theta_{\rm E}}}\right)^{2-\gamma}},
\end{align}
where
\begin{align}
F=\left[\displaystyle{\frac{\Gamma[(\xi-1)/2]}{\Gamma(\xi/2)}}-\beta\displaystyle{\frac{\Gamma[(\xi+1)/2]}{\Gamma[(\xi+2)/2)]}}\right]
\displaystyle{\frac{\Gamma(\gamma/2)\Gamma(\delta/2)}{\Gamma[(\gamma-1)/2]\Gamma[(\delta-1)/2]}}.
\end{align}
\end{small}
In the case of $\gamma=\delta=2$ and $\beta=0$, the mass model is simplified to the singular isothermal sphere model, then
\begin{align}
\label{SGL-VD}
\sigma_0=\sqrt{\displaystyle{\frac{c^2}{4\pi}}\displaystyle{\frac{D_{\rm s}}{D_{\rm ls}}}\theta_{\rm E}}.
\end{align}
{Once $\theta_{\rm E}$ is obtained, we can calculate $\sigma_0$ under a specific cosmological model. By comparing the calculated $\sigma_0$ with the observed one, we can put constrains on the assumed model by the distance ratio $D_{\rm s}/D_{\rm ls}$. Note that $H_0$ exists in both $D_{\rm s}$ and $D_{\rm ls}$ expressions, but it cancels out in the distance ratio, so the VD observations cannot place constraints on $H_0$.} We use the code provided in Ref.~\citep{Collett:2015roa} to simulate 8000 VD events of future LSST. The mock data include the lens redshifts $z_{\rm l}$, the source redshifts $z_{\rm s}$, the Einstein angles $\theta_{\rm E}$, and the velocity dispersions $\sigma_0$. We simply assume that the relative errors of $\sigma_0$ are $5\%$ \citep{Qi:2018atg}. The
$\chi^2$ function of VD can be written as
\begin{align}
\chi_{\rm VD}^2(\boldsymbol{\xi}) &= \sum_{i=1}^{8000}\left(\displaystyle{\frac{\sigma_{0,i}^{\rm th}(\boldsymbol{\xi})-\sigma_{0,i}^{\rm obs}}{\sigma(\sigma_{0,i})}}\right)^2,
\end{align}
where $\boldsymbol{\xi}$ denotes a set of cosmological parameters.

\begin{figure}[!htbp]
\includegraphics[scale=0.53]{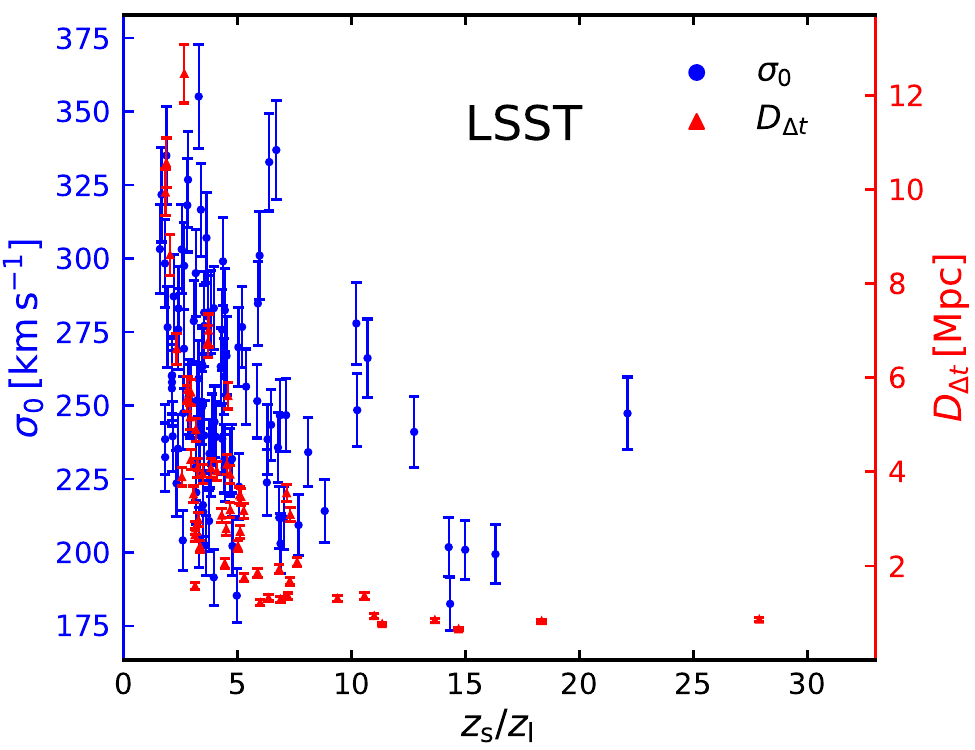}
\centering
\caption{The simulated TD and VD data based on LSST. Note that we actually simulate 8000 VD events, but only 100 of them are shown here.}
\label{sgl}
\end{figure}

The gravitational lens time-delay method is another cosmological application of SGL systems (see e.g. Refs. \cite{Birrer:2018vtm, Birrer:2020tax,Qi:2022sxm,Cao:2021zpf}). If the source has flux variations, time delays between multiple images can be measured by monitoring the lens. The time delay between images $i$ and $j$ is given by
\begin{align}
\Delta t_{ij}=\displaystyle{\frac{D_{\Delta t}}{c}}\left[\displaystyle{\frac{({\boldsymbol{\theta}}_i-\boldsymbol{\beta})^2}{2}}-\psi({\boldsymbol{\theta}}_i)-\displaystyle{\frac{({\boldsymbol{\theta}}_j-\boldsymbol{\beta})^2}{2}}+\psi({\boldsymbol{\theta}}_j)\right],
\end{align}
where $D_{\Delta t}$ is the time-delay distance, calculated by
\begin{align}
\label{SGL-TD}
D_{\Delta t}\equiv(1+z_{\rm l})\displaystyle{\frac{D_{\rm l}D_{\rm s}}{D_{\rm ls}}},
\end{align}
$\boldsymbol{\beta}$ is the source position, $\boldsymbol{\theta}_i$ is the image position, and $\psi$ is the lensing potential. By measuring $\boldsymbol{\beta}$, $\boldsymbol{\theta}_i$, $\psi(\boldsymbol{\theta}_i)$, and $\Delta t_{ij}$, one can obtain $D_{\rm l}D_{\rm s}/D_{\rm ls}$, which is closely related to cosmology. {Note that the ratio of three distances will retain an $H_0$, so the TD observation is capable of constraining the Hubble constant.} In this paper, we assume that 55 TD events can be measured, and the redshifts of the sources and lenses are taken from the VD data simulated above. We calculate the time-delay distances $D_{\Delta t}$ in the $\Lambda$CDM model and also take $5\%$ relative errors for them. The simulated VD and TD events are shown in Fig.~\ref{sgl}.

The $\chi^2$ functions of TD and SGL are then given by
\begin{align}
\chi_{\rm TD}^2(\boldsymbol{\xi}) &= \sum_{i=1}^{55}\left(\displaystyle{\frac{D_{\Delta t,i}^{\rm th}(\boldsymbol{\xi})-D_{\Delta t,i}^{\rm obs}}{\sigma (D_{\Delta t,i})}}\right)^2,\nonumber \\
\chi_{\rm SGL}^2&=\chi_{\rm VD}^2+\chi_{\rm TD}^2.
\end{align}

In this work, the four late-universe cosmological probes are uncorrelated. Taking all of these into account, the total $\chi^2$ function of the four probes can be written as
\begin{align}
\chi_{\rm tot}^2=\chi_{\rm 21\,cm\,IM}^2 +\chi_{\rm FRB}^2 +\chi_{\rm GW}^2 +\chi_{\rm SGL}^2.
\end{align}

In this paper, we focus on the synergy of the four late-universe probes. We mainly wish to investigate how they can constrain the late-universe physics (such as dark energy and the Hubble constant), and thus we do not concern the primordial-universe parameters (such as $n_{\rm s}$). When generating the mock data of these observations, we do not consider the fluctuations in central values of the mock data. The reasons are as follows: (i) In a forecast study, only the constraint errors of cosmological parameters are important, but their central values are not worth concerning. (ii) Since we wish to combine the four cosmological probes, it is necessary to try to avoid the potential tensions between them. (iii) In order to clearly show how the cosmological parameter degeneracies are broken by the synergy of the probes, the central values in the contour plots are better to be well concordant.

\section{Results and discussions}\label{sec3}
\begin{figure*}[!htbp]
\includegraphics[scale=0.37]{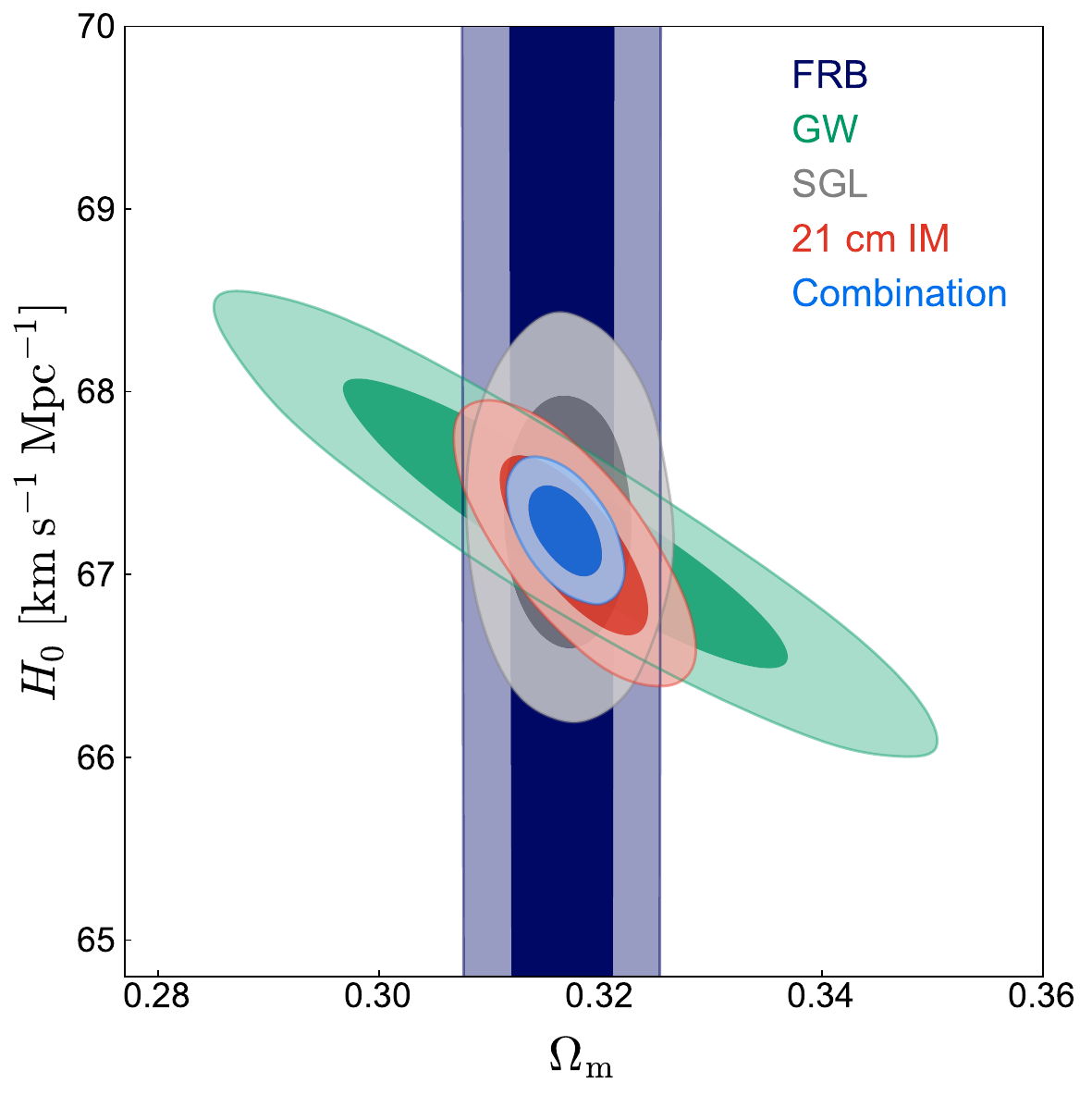}
\includegraphics[scale=0.37]{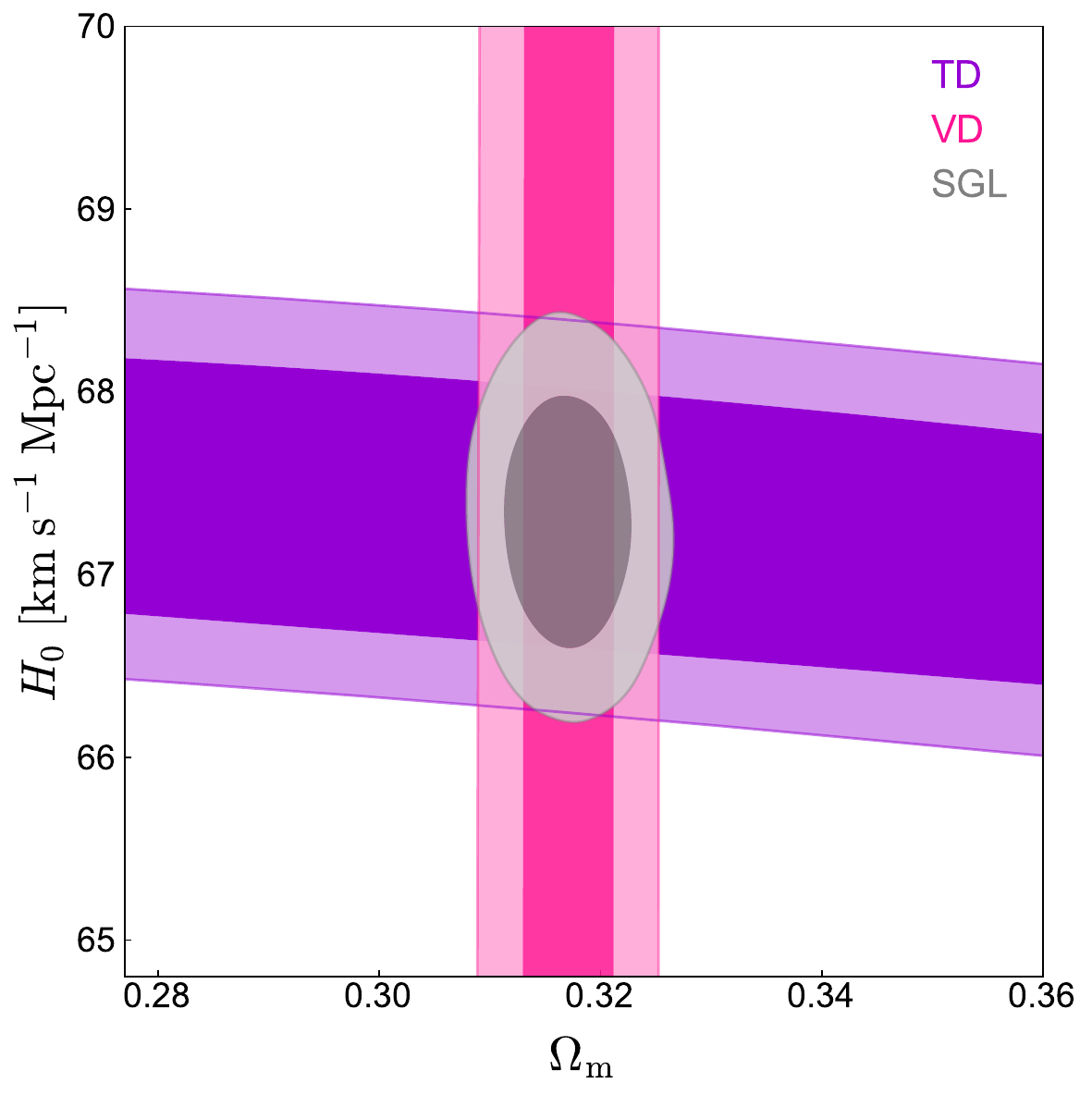}
\caption{Left panel: Constraints (68.3\% and 95.4\% confidence level) on the $\Lambda$CDM model by using the FRB, GW, SGL, 21\,cm\,IM, and 21\,cm\,IM+FRB+GW+SGL data. Right panel: Constraints on the $\Lambda$CDM model by using the TD, VD, and SGL data.}
\label{fig:LCDM}
\end{figure*}

\begin{figure*}[!htbp]
\centering
\includegraphics[scale=0.355]{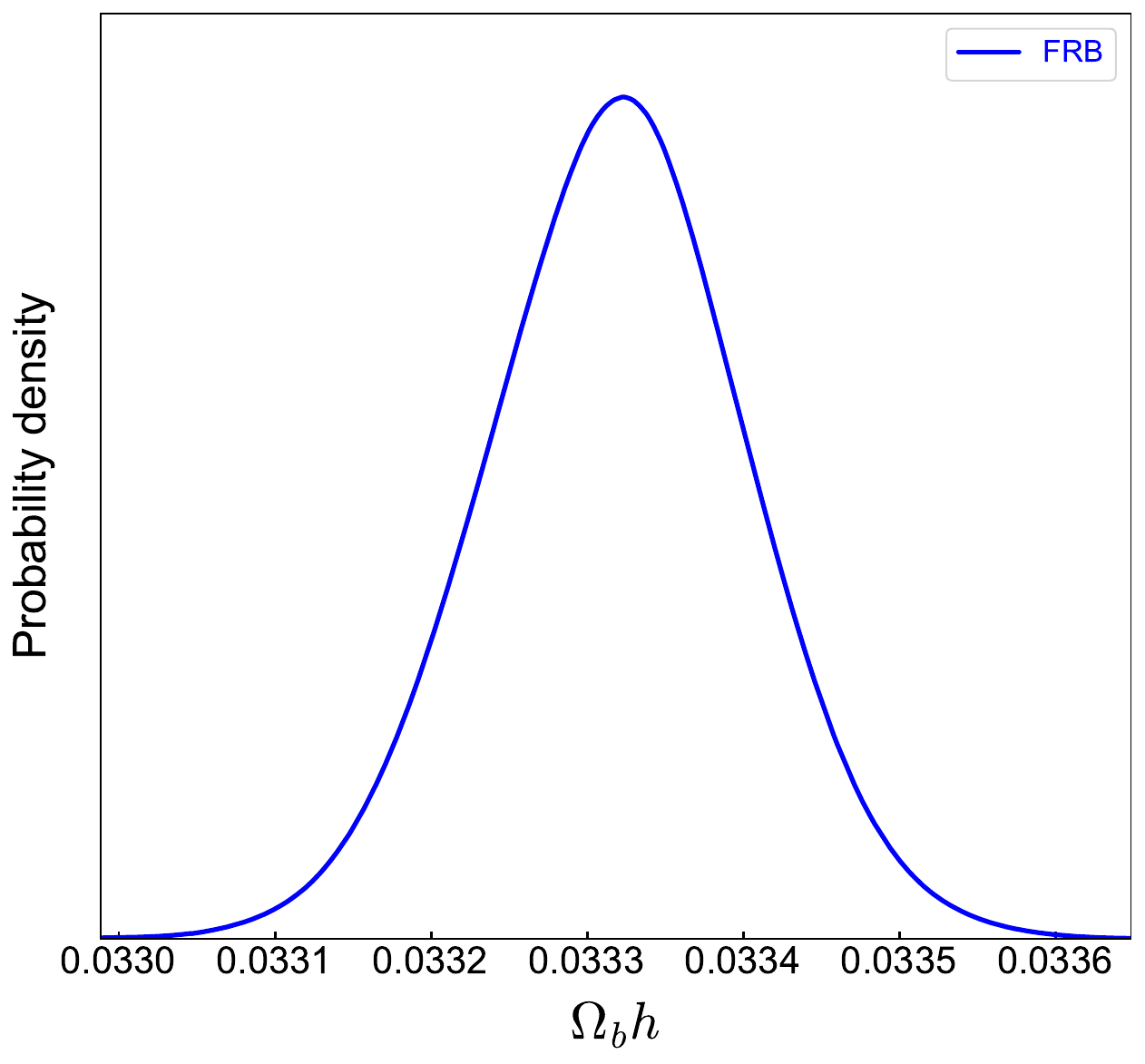}
\includegraphics[scale=0.37]{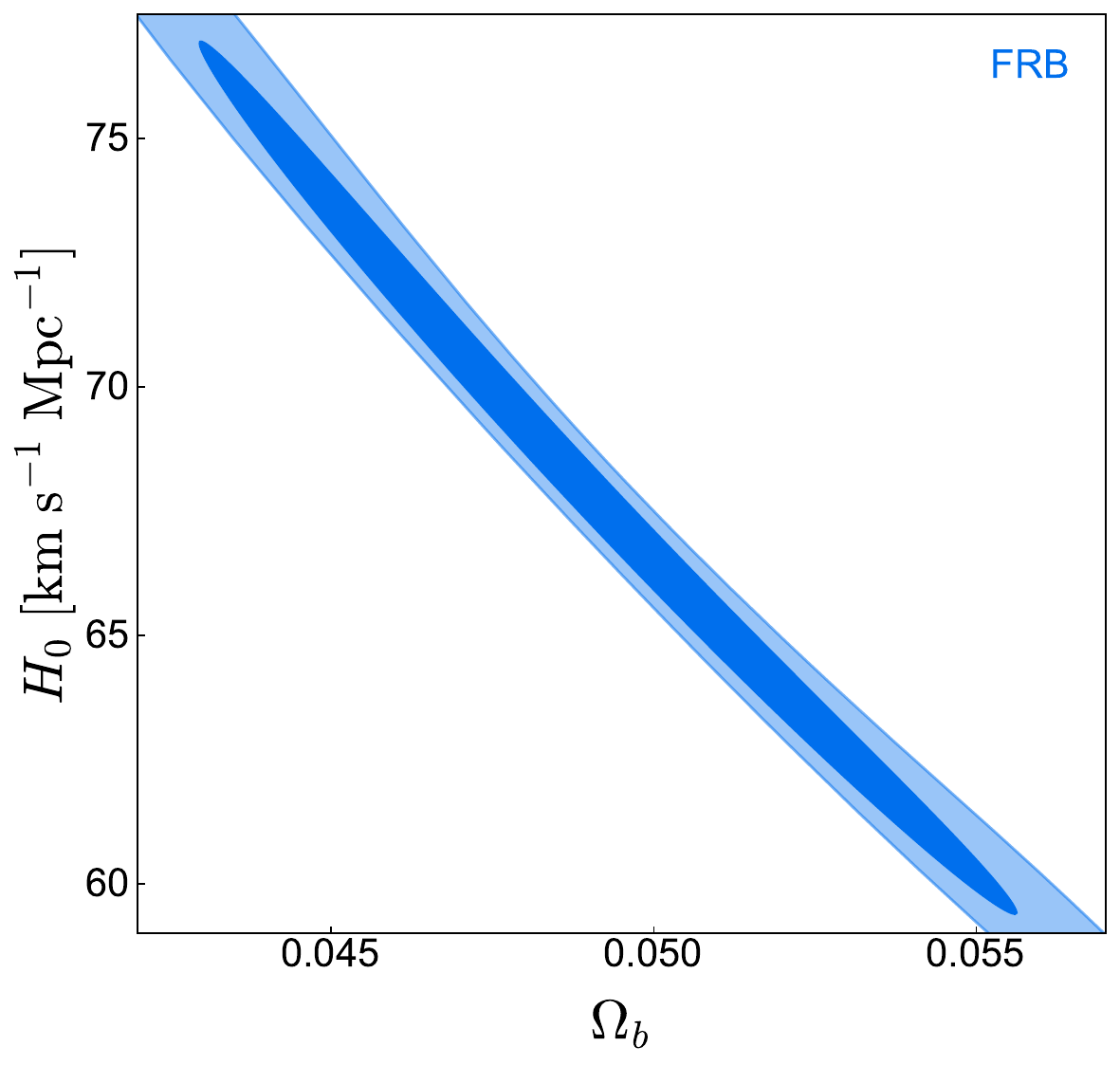}
\caption{Constraints (68.3\% and 95.4\% confidence level) on the $\Lambda$CDM model by using the FRB data. Left panel: the constraint on the parameter combination $\Omega_{\rm b}h$. Right panel: the $\Omega_{\rm b}$--$H_0$ degeneracy.}
\label{fig:FRB-compare}
\end{figure*}

\begin{figure*}[!htbp]
\includegraphics[scale=0.53]{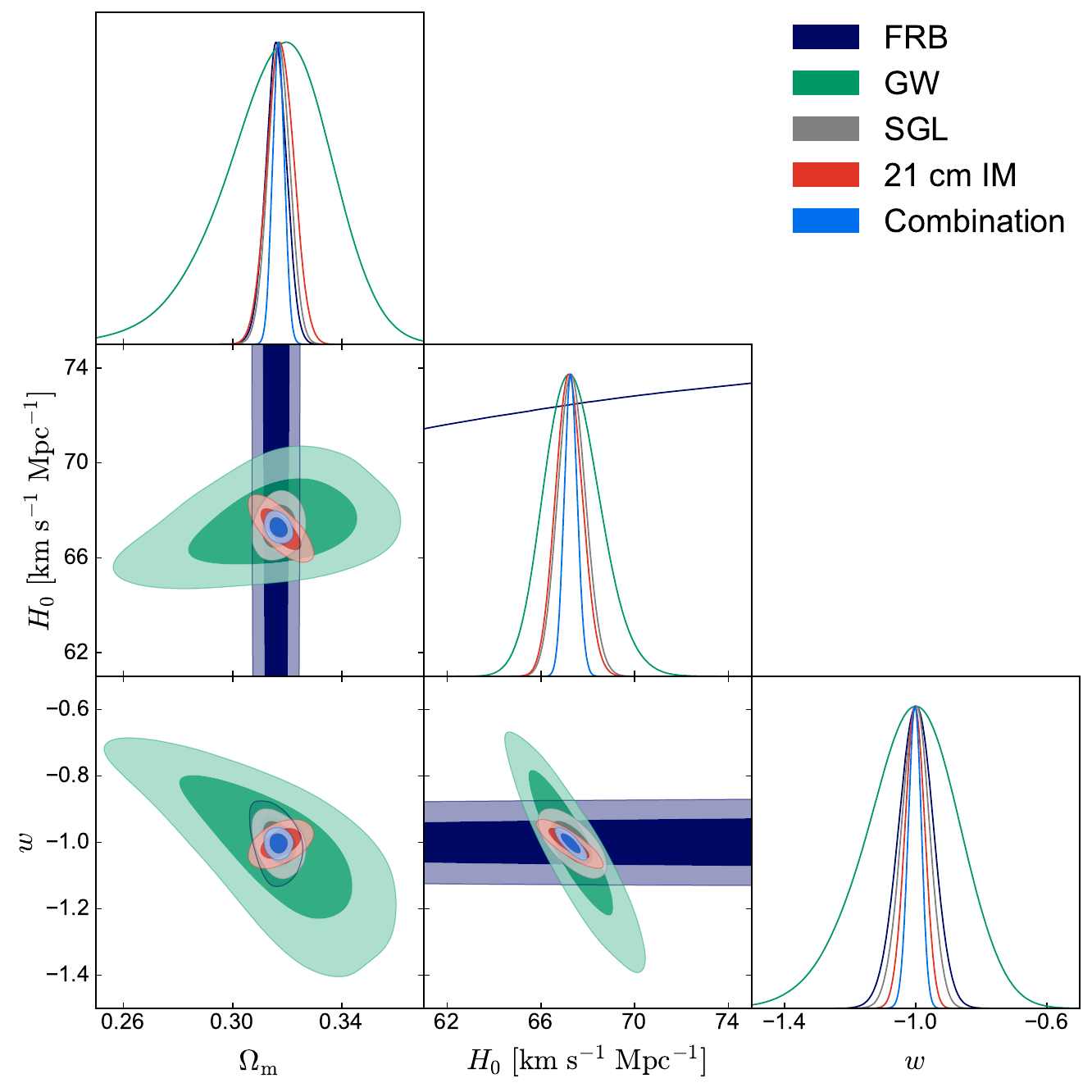}
\centering
\caption{Constraints (68.3\% and 95.4\% confidence level) on the $w$CDM model by using the  FRB, 21\,cm\,IM, GW, SGL, and 21\,cm\,IM+FRB+GW+SGL data.}
\label{fig:wCDM}
\end{figure*}

\begin{figure}[!htbp]
\includegraphics[scale=0.37]{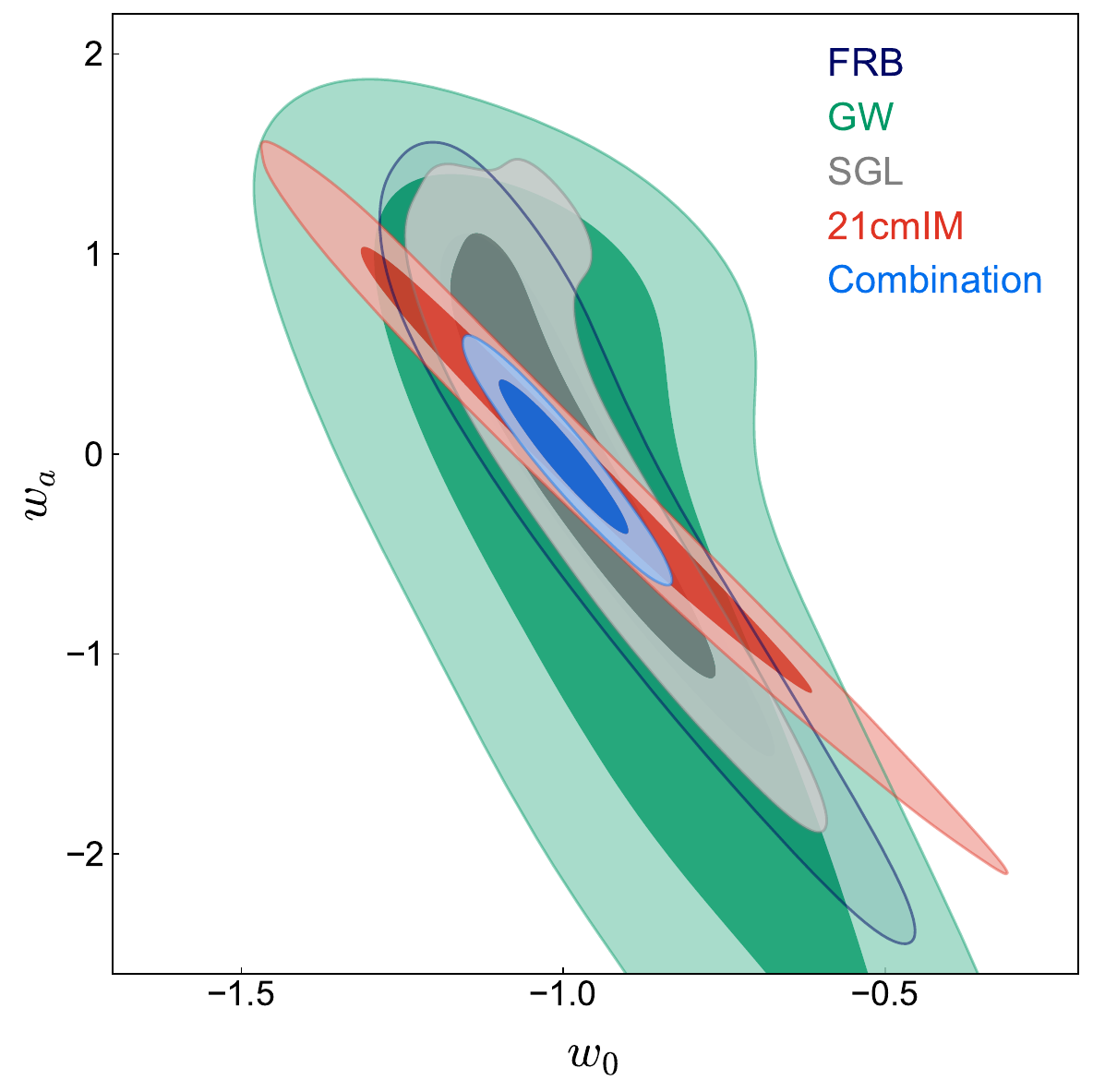}
\centering
\caption{Constraints (68.3\% and 95.4\% confidence level) on the $w_0w_a$CDM model by using the  FRB, GW, SGL, 21\,cm\,IM, and 21\,cm\,IM+FRB+GW+SGL data.}
\label{fig:CPL}
\end{figure}

In this section, we report the constraint results from 21\,cm\,IM, FRB, GW, SGL, and the combination of them. Here we consider only the three most typical cosmological models of dark energy: (\romannumeral1) $\Lambda$CDM model---the standard cosmological model with $w(z)=-1$; (\romannumeral2) $w$CDM model---the simplest dynamical dark energy model with a constant equation of state (EoS) $w(z)=w$; (\romannumeral3) $w_0w_a$CDM model---the dynamical dark energy model with a parameterized EoS $w(z)=w_0+w_a z/(1+z)$ \citep{Chevallier:2000qy,Linder:2002et}. The cosmological parameters we sample include $H_0$, $\Omega_{\rm m}$, $\Omega_{\rm b}h^2$, $\sigma_8$, $w$, $w_0$, and $w_a$, and we take flat priors for them. The $1\sigma$ and $2\sigma$ posterior distribution contours for various model parameters are shown in Figs.~\ref{fig:LCDM}--\ref{fig:CPL}, and the $1\sigma$ errors for the marginalized parameter constraints are summarized in Table~\ref{tab:result}. In the following discussions, we use $\sigma(\xi)$ and $\varepsilon(\xi)=\sigma(\xi)/\xi$ to represent the absolute and relative errors of the cosmological parameter $\xi$, respectively.

\begin{table*}[!htb]
\caption{The absolute (1$\sigma$) and relative errors of the cosmological parameters in the $\Lambda$CDM, $w$CDM and $w_0w_a$CDM models using the 21\,cm\,IM, FRB, GW, VD, TD, SGL, and 21\,cm\,IM+FRB+GW+SGL data. Here $H_0$ is in units of $\rm km\ s^{-1}\ Mpc^{-1}$. Note that $\sigma(\xi)$ and $\varepsilon(\xi)=\sigma(\xi)/\xi$ represent the absolute and relative errors of the cosmological parameter $\xi$, respectively.}
\label{tab:result}
\setlength{\tabcolsep}{2mm}
\renewcommand{\arraystretch}{1.2}
\begin{center}{\centerline{
\begin{tabular}{ccm{1.3cm}<{\centering}m{1cm}<{\centering}m{1cm}<{\centering}m{1cm}<{\centering}m{1cm}<{\centering}m{1cm}<{\centering}m{3.8cm}<{\centering}m{2cm}<{\centering}}
\hline
Model       & Error                             &21\,cm\,IM         &FRB        &GW         &VD         &TD         &SGL        &21\,cm\,IM+FRB+GW+SGL  \\ \hline
 \multirow{4}{*}{$\Lambda$CDM}
            &$\sigma(\Omega_{\rm m})$           &$0.0044$           &$0.0036$   &$0.013$    &$0.0038$   &$-$       &$0.0038$   &$0.0022$               \\
            &$\sigma(H_0)$                      &$0.32$             &$-$        &$0.52$     &$-$        &$0.74$     &$0.46$     &$0.16$                 \\
            &$\varepsilon(\Omega_{\rm m})$      &$1.4\%$            &$1.1\%$    &$4.1\%$    &$1.2\%$    &$-$       &$1.2\%$    &$0.7\%$                \\
            &$\varepsilon(H_0)$                 &$0.5\%$            &$-$        &$0.8\%$    &$-$        &$1.1\%$    &$0.7\%$    &$0.2\%$                \\  \hline
 \multirow{6}{*}{$w$CDM}
            &$\sigma(\Omega_{\rm m})$           &$0.0049$           &$0.0039$   &$0.019$    &$0.0040$   &$0.15$     &$0.0040$   &$0.0022$               \\
            &$\sigma(H_0)$                      &$0.58$             &$-$        &$1.2$      &$-$        &$2.3$      &$0.60$     &$0.28$                 \\
            &$\sigma(w)$                        &$0.030$            &$0.053$    &$0.14$     &$0.046$    &$0.48$     &$0.043$    &$0.020$                \\
            &$\varepsilon(\Omega_{\rm m})$      &$1.5\%$            &$1.2\%$    &$6.0\%$    &$1.3\%$    &$47\%$     &$1.3\%$    &$0.7\%$                \\
            &$\varepsilon(H_0)$                 &$0.9\%$            &$-$        &$1.8\%$    &$-$        &$3.4\%$    &$0.9\%$    &$0.4\%$                \\
            &$\varepsilon(w)$                   &$3.0\%$            &$5.3\%$    &$14\%$     &$4.6\%$    &$48\%$     &$4.3\%$    &$2.0\%$                \\ \hline
 \multirow{7}{*}{$w_0w_a$CDM}
            &$\sigma(\Omega_{\rm m})$           &$0.030$            &$0.035$    &$0.045$    &$0.030$    &$0.15$     &$0.029$    &$0.0092$               \\
            &$\sigma(H_0)$                      &$2.4$              &$-$        &$1.6$      &$-$        &$2.5$      &$1.2$      &$0.61$                 \\
            &$\sigma(w_0)$                      &$0.23$             &$0.16$     &$0.22$     &$0.13$     &$0.68$     &$0.12$     &$0.066$                \\
            &$\sigma(w_a)$                      &$0.74$             &$0.83$     &$1.3$      &$0.76$     &$1.8$      &$0.75$     &$0.25$                 \\
            &$\varepsilon(\Omega_{\rm m})$      &$9.5\%$            &$11\%$     &$14\%$     &$9.5\%$    &$47\%$     &$9.2\%$    &$2.9\%$                \\
            &$\varepsilon(H_0)$                 &$3.6\%$            &$-$        &$2.4\%$    &$-$        &$3.7\%$    &$1.8\%$    &$0.9\%$                \\
            &$\varepsilon(w_0)$                 &$23\%$             &$16\%$     &$22\%$     &$13\%$     &$68\%$     &$12\%$     &$6.6\%$                \\  \hline
\end{tabular}}}
\end{center}
\end{table*}

In the left panel of Fig.~\ref{fig:LCDM}, we show the constraints on $\Lambda$CDM in the $\Omega_{\rm m}$--$H_0$ plane. It is obvious that the Hubble constant $H_0$ cannot be well constrained by FRB alone, since the dispersion measure from the intergalactic medium ${\rm DM}_{\rm IGM}$ is proportional to $H_0\Omega_{\rm b}$ [see Eq.~(\ref{FRB-DM-IGM})]. In contrast, 21\,cm\,IM, GW, and SGL can provide the small constraint errors of 0.32, 0.52, and 0.46 for $H_0$, respectively, all meeting the standard $\varepsilon(H_0)<1\%$.

It should be pointed out that the 21\,cm\,IM alone cannot constrain $H_0$ but only $H_0 r_{\rm d}$ (with $r_{\rm d}$ the sound horizon at the drag epoch where baryons decouple from photons). The $H_0$ constraint from BAO actually needs the addition of other observations such as CMB or big bang nucleosynthesis (BBN) data helping break the $H_0$--$r_{\rm d}$ degeneracy \cite{Addison:2017fdm}. In the forecast, we have chosen the Planck best-fit $\Lambda$CDM model as a fiducial model to generate the mock data, which is equivalent to inputting the Planck best-fit $r_{\rm d}$ into the 21\,cm\,IM data to break the $H_0$--$r_{\rm d}$ degeneracy in BAO measurement, therefore the $H_0$ constraint from 21\,cm\,IM here actually includes some contribution from CMB. Of course, the effect of CMB in this case is mainly on the central value of $H_0$, but the error of $H_0$ is slightly affected. In the future, when the actual observational data of 21 cm IM could be used in addressing the Hubble tension, any connection with CMB should be avoided.

In Fig.~\ref{fig:FRB-compare}, we show the tight constraint on $\Omega_{\rm b}h$ (left panel) and the strong degeneracy between $\Omega_{\rm b}$ and $H_0$ (right panel) in $\Lambda$CDM from the FRB mock data. Therefore, using the localized FRBs to determine the baryon density needs to assume a value of the Hubble constant \cite{Macquart:2020lln}. Likewise, the determination of the Hubble constant using the FRB observation also needs the help of the baryon density constraints from other observations \cite{Zhao:2022yiv,Wu:2021jyk, Hagstotz:2021jzu, Liu:2022bmn, James:2022dcx}.

Although FRB alone cannot effectively constrain $H_0$, it gives a tight constraint on $\Omega_{\rm m}$, $\sigma(\Omega_{\rm m})=0.0036$ and $\varepsilon(\Omega_{\rm m})=1.1\%$, which is slightly better than those of 21\,cm\,IM, GW, and SGL. Therefore, combining FRB with 21\,cm\,IM, GW, or SGL can effectively constrain $\Omega_{\rm m}$ and $H_0$ at the same time. In addition, 21\,cm\,IM, GW, and SGL have obviously different parameter dependencies, so any combination of them can break the degeneracies and thus improve the constraint precision. In general, any combination of the four probes is meaningful and worth expecting. One may find that a large fraction of the constraining power comes from 21\,cm\,IM. As mentioned earlier, the 21\,cm\,IM technique can measure the LSS of the universe without having to resolve individual galaxies, which makes it much faster to survey large volumes than traditional galaxy redshift surveys. It is worth mentioning that future galaxy redshift surveys are still important, although they are more time-consuming. In Ref.~\citep{Bull:2014rha}, the comparison between the DETF Stage IV galaxy surveys, such as Euclid \citep{Amendola:2016saw} and LSST \citep{LSST:2008ijt}, and the future 21\,cm\,IM experiments, has been made, and it was found that the future 21\,cm\,IM experiments would have a comparable capability in constraining cosmological parameters.

Note that the performance of SGL in cosmological constraints is actually the result of the combination of VD and TD. We know that VD and TD relates to cosmology by the angular diameter distance ratios $D_{\rm s}/D_{\rm ls}$ and $D_{\rm l}D_{\rm s}/D_{\rm ls}$, respectively [see Eqs.~(\ref{SGL-VD}) and (\ref{SGL-TD})]. $H_0$ is cancelled out in the former but retained in the latter, so VD is insensitive to $H_0$ but TD is very sensitive to $H_0$. In the right panel of Fig.~\ref{fig:LCDM}, we show the constraints on $\Lambda$CDM by using the VD, TD, and SGL (i.e., VD+TD) data. It can be seen that the contours from VD and TD are almost orthogonal, so VD+TD can thoroughly break the parameter degeneracies inherent to VD and TD alone. As a result, the SGL data can provide the tight constraints, $\sigma(\Omega_{\rm m})=0.0038$, $\sigma(H_0)=0.46\ \rm km\ s^{-1}\ Mpc^{-1}$, $\varepsilon(\Omega_{\rm m})=1.2\%$, and $\varepsilon(H_0)=0.7\%$. It is worth noting that FRB and VD behave very similarly in constraining cosmological parameters in $\Lambda$CDM, both of which can tightly constrain $\Omega_{\rm m}$ but not $H_0$. Excitingly, the joint 21\,cm\,IM+FRB+GW+SGL data gives $\sigma(\Omega_{\rm m})=0.0022$, $\sigma(H_0)=0.16\ \rm km\ s^{-1}\ Mpc^{-1}$, $\varepsilon(\Omega_{\rm m})=0.7\%$, and $\varepsilon(H_0)=0.2\%$, which has achieved the standard of precision cosmology, i.e., the precision of parameters is better than 1\%.

In Fig.~\ref{fig:wCDM}, we show the $1\sigma$ and $2\sigma$ posterior distribution contours for the $w$CDM model. We can see that the four probes have different parameter dependencies and thus the combination of them could break the degeneracies. Concretely, the joint data provide $\sigma(w)=0.020$, which is 35\% better than the result of $\sigma(w)=0.031$ obtained by the CMB+BAO+SN data \citep{Aghanim:2018eyx}. Moreover, the constraint precision of $\Omega_{\rm m}$ and $H_0$ is still better than 1\%. Note that CMB+BAO+SN is the combination of early and late-universe observations, and here we use only the combination of late-universe probes.

In Fig.~\ref{fig:CPL}, we show the constraints on $w_0w_a$CDM in the $w_0$--$w_a$ plane that we are most interested in. It can be seen that the performance of 21\,cm\,IM in constraining the parameterized dynamical dark energy model is far inferior to that in constraining the $\Lambda$CDM and $w$CDM models. This is because the surveys in the dark energy-dominated era of the universe help to better constrain the dynamical dark energy model, while HIRAX is designed to mainly cover the matter-dominated era of the universe ($0.8<z<2.5$) \citep{Wu:2021vfz}. However, 21\,cm\,IM has a different $w_0$--$w_a$ degeneracy orientation from the other three probes. The joint constraint provides $\sigma(w_0)=0.066$ and $\sigma(w_a)=0.25$, which are 18\% and 14\% better than the constraint results of $\sigma(w_0)=0.080$ and $\sigma(w_a)=0.29$ achieved by the CMB+BAO+SN data, respectively \citep{Aghanim:2018eyx}.

It is known that the $H_0$ tension between the Cepheid-supernova distance ladder measurement \citep{Riess:2020fzl} and the Planck CMB inference \citep{Aghanim:2018eyx} has now reached $4.2\sigma$. To solve the $H_0$ tension, on one hand, it is important to develop the new late-universe cosmological probes independent of the distance ladder to precisely measure the Hubble constant, and on the other hand, from the point of view of searching for new physics in cosmology, it is also of great importance to use new late-universe probes to precisely constrain new-physics effects and the related parameters. Therefore, from the both points of view, it is fairly necessary to vigorously develop new late-universe cosmological probes in the next decades. Although the CMB measurements initiated the era of precision cosmology, they can only precisely constrain the cosmological parameters in the $\Lambda$CDM model. Since the CMB observation is an early universe probe, it cannot effectively constrain the late-universe physical effects, in particular, the CMB data can only provide rather poor constraints on the EoS of dark energy. Hence, in order to precisely constrain the Hubble constant and the EoS of dark energy at the same time, in this work we propose that in the next decades we need to forge precise late-universe cosmological probes, in particular, 21\,cm\,IM, FRB, GW, and SGL, and consider the synergy of them in exploring the nature of dark energy and solving the Hubble tension.

\section{Conclusion}\label{sec4}

In the next decades, it is necessary to develop new late-universe cosmological probes to precisely measure the Hubble constant and the EoS of dark energy at the same time. In this work, we show that the four typical late-universe cosmological probes, 21\,cm\,IM, FRB, GW standard siren, and SGL, will play an important role in cosmology in the near future. We investigate the capability of their combination to constrain cosmological parameters. Here, the 21\,cm\,IM, FRB, GW, and SGL data are simulated based on the hypothetical observations of HIRAX, SKA, ET, and LSST, respectively.

We find that 21\,cm\,IM, GW, and SGL all can constrain the Hubble constant $H_0$ to the precision better than $1\%$ in the $\Lambda$CDM model, so they will play an important role in solving the $H_0$ tension. Importantly, 21\,cm\,IM, FRB, GW, and SGL have different parameter dependencies and thus any combination of them could effectively break the degeneracies. It should be pointed out that SGL is composed of VD and TD in this work. The parameter degeneracy orientations of VD and TD are almost orthogonal in cosmological constraints, so VD+TD can thoroughly break the degeneracies inherent to VD and TD alone. The SGL data can offer the tight constraints, $\sigma(\Omega_{\rm m})=0.0038$ and $\sigma(H_0)=0.46\ \rm km\ s^{-1}\ Mpc^{-1}$, mainly for this reason. In addition, FRB and VD behave very similarly in constraining cosmological parameters in $\Lambda$CDM, both of which can tightly constrain $\Omega_{\rm m}$ but not $H_0$.

The joint 21\,cm\,IM+FRB+GW+SGL data could provide the constraint errors of $\sigma(\Omega_{\rm m})=0.0022$ and $\sigma(H_0)=0.16\ \rm km\ s^{-1}\ Mpc^{-1}$ in the $\Lambda$CDM model, which has achieved the standard of precision cosmology, i.e., the precision of parameters is better than 1\%. Moreover, the joint data can tightly constrain the dynamical dark energy EoS parameters. It offers $\sigma(w)=0.020$ in the $w$CDM model, and $\sigma(w_0)=0.066$ and $\sigma(w_a)=0.25$ in the $w_0w_a$CDM model, which are better than the constraint results achieved by the CMB+BAO+SN data \citep{Aghanim:2018eyx}. Our results are sufficient to show that the synergy of the four late-universe cosmological probes has magnificent prospects in cosmological studies.

\begin{acknowledgments}
We are grateful to Ji-Guo Zhang, Ze-Wei Zhao, Jing-Zhao Qi, Yichao Li, Wei-Hong Hu, Yu Cui, and Jing-Fei Zhang for the fruitful discussions.
This work was supported by the National SKA Program of China (Grants Nos. 2022SKA0110200 and 2022SKA0110203) and the National Natural Science Foundation of China (Grants Nos. 11975072, 11835009, and 11875102).
\end{acknowledgments}

\bibliography{probes}{}
\bibliographystyle{JHEP}

\end{document}